\documentclass[twocolumn,trackchanges]{aastex63}

\usepackage{amsmath}
\usepackage{graphicx}
\usepackage{appendix}
\usepackage{comment}
\usepackage{tabularx}
\usepackage{booktabs}
\usepackage{comment}
\usepackage{xcolor}
\usepackage{lineno}

\newcommand{\isofont}[1]{{\mathrm{#1}}}
\newcommand{\isomass}[1]{{\ensuremath{\isofont{^{#1}}}}}
\newcommand{\isocharge}[1]{{\ensuremath{\isofont{_{#1}}}}}
\newcommand{\isotope}[3]{{\ensuremath{\isocharge{#1}\isomass{#2}\isofont{#3}}}}

\newcommand{\I}[2]{{\isotope{}{#1}{#2}}}

\newcommand{\Ep}[1]{{\ensuremath{10^{#1}}}}
\newcommand{\E}[1]{{\ensuremath{\powersep\Ep{#1}}}}

\newcommand{\powersep}{\times}

\newcommand{\unitstyle}[1]{\ensuremath{\mathrm{#1}}}

\newcommand{\Kelvin}{\unitstyle{K}}
\newcommand{\K}{\Kelvin}  

\newcommand{\code}[1]{\texttt{#1}}
\newcommand{\mesa}{\code{MESA}}
\newcommand{\MESA}{\mesa}

\input{vectors.tex}

\newcommand{\D}{{\mathrm d}}

\newcommand{\Fgamma}{\ensuremath{F_\gamma}}     

\def\partf#1#2{{{\partial{#1}}/{\partial{#2}}}} 

\newlength{\apjcolwidth}
\newcommand{\tightitems}{
  \setlength{\topsep}{0pt}
  \setlength{\itemsep}{-1pt}
  \setlength{\parsep}{0pt}
  \setlength{\parskip}{0pt}
}

\begin{document}
\title{An Expanded Set of Los Alamos OPLIB Tables in MESA: \\ Type-1 Rosseland-mean Opacities and Solar Models}

\shorttitle{OPLIB Tables in MESA: I. Type-1 Rosseland-mean Opacities and Solar Models}
\shortauthors{Farag et al.} 

\author[0000-0002-5794-4286]{Ebraheem Farag}
\affiliation{School of Earth and Space Exploration, Arizona State University, Tempe, AZ 85287, USA}
\affiliation{Center for Theoretical Astrophysics, Los Alamos National Laboratory, Los Alamos, NM 87545}
\affiliation{Department of Astronomy, Yale University, New Haven, CT 06511}

\author[0000-0003-1087-2964]{Christopher J. Fontes}
\affiliation{Center for Theoretical Astrophysics, Los Alamos National Laboratory, Los Alamos, NM 87545}

\author[0000-0002-0474-159X]{F.X.~Timmes}
\affiliation{School of Earth and Space Exploration, Arizona State University, Tempe, AZ 85287, USA}

\author[0000-0003-4456-4863]{Earl P. Bellinger}
\affiliation{Department of Astronomy, Yale University, New Haven, CT 06511}
\affiliation{Max-Planck-Institut f\"{u}r Astrophysik, Karl-Schwarzschild-Stra{\ss}e 1, D-85741 Garching, Germany}
\affiliation{Stellar Astrophysics Centre, Aarhus University, Aarhus, Denmark}

\author[0000-0003-1291-1533]{Joyce A. Guzik}
\affiliation{Center for Theoretical Astrophysics, Los Alamos National Laboratory, Los Alamos, NM 87545}

\author[0000-0002-4791-6724]{Evan B. Bauer}
\affiliation{Center for Astrophysics $\vert$ Harvard \& Smithsonian, 60 Garden Street, Cambridge, MA 02138, USA}

\author[0000-0002-7208-7681]{Suzannah R. Wood}
\affiliation{Center for Theoretical Astrophysics, Los Alamos National Laboratory, Los Alamos, NM 87545}

\author[0000-0002-5539-9034]{Katie Mussack}
\affiliation{Center for Theoretical Astrophysics, Los Alamos National Laboratory, Los Alamos, NM 87545}

\author[0000-0002-7936-4231]{Peter Hakel}
\affiliation{Los Alamos National Laboratory, Los Alamos, NM 87545}

\author[0000-0003-1045-3858]{James Colgan}
\affiliation{Los Alamos National Laboratory, Los Alamos, NM 87545}

\author[0000-0002-2319-5934]{David P. Kilcrease}
\affiliation{Los Alamos National Laboratory, Los Alamos, NM 87545}

\author[0000-0002-0312-7694]{Manolo E. Sherrill}
\affiliation{Los Alamos National Laboratory, Los Alamos, NM 87545}

\author[0009-0002-2268-7352]{Tryston C. Raecke}
\affiliation{Los Alamos National Laboratory, Los Alamos, NM 87545}

\author[0000-0002-5107-8639]{Morgan T. Chidester}
\affiliation{School of Earth and Space Exploration, Arizona State University, Tempe, AZ 85287, USA}

\correspondingauthor{Ebraheem Farag}
\email{ekfarag@asu.edu}

\begin{abstract}


We present a set of 1194 Type-1 Rosseland-mean opacity tables for four different metallicity mixtures. These new Los Alamos OPLIB atomic radiative opacity tables are an order of magnitude larger in number than any previous opacity table release, and span regimes where previous opacity tables have not existed. For example, the new set of opacity tables expands the metallicity range to $Z$\,=\,10$^{-6}$ to $Z$\,=\,0.2 which  allows improved accuracy of opacities at low and high metallicity, increases the table density in the metallicity range $Z$\,=\,10$^{-4}$ to $Z$\,=\,0.1 to enhance the accuracy of opacities drawn from interpolations across neighboring metallicities, and adds entries for hydrogen mass fractions between $X$\,=\,0 and $X$\,=\,0.1 including $X$\,=\,$10^{-2}, 10^{-3}, 10^{-4}, 10^{-5}, 10^{-6}$ that can improve stellar models of hydrogen deficient stars. We implement these new OPLIB radiative opacity tables in \MESA, and find that calibrated solar models agree broadly with previously published helioseismic and solar neutrino results. We find differences between using the new 1194 OPLIB opacity tables and the 126 OPAL opacity tables range from $\approx$\,20--80\% across individual chemical mixtures, up to $\approx$\,8\% and $\approx$\,15\% at the bottom and top of the solar convection zone respectively, and $\approx$\,7\% in the solar core. We also find differences between standard solar models using different opacity table sources that are on par with altering the initial abundance mixture. We conclude that this new, open-access set of OPLIB opacity tables does not solve the solar modeling problem, and suggest the investigation of physical mechanisms other than the atomic radiative opacity.

\end{abstract}


\section{Introduction} \label{s.intro}

The radiative opacity $\kappa$, a cross-section per unit mass usually expressed in cm$^2$ g$^{-1}$, describes the absorption and scattering of photons in a medium. It is a crucial component of stellar physics that, in the interior, relates the diffusive transport of the photon energy flux \Fgamma\ to spatial gradients in the energy density of the radiation field: \Fgamma\,$\propto$\,(1/$\kappa$) \ d($T^4$)/d$r$,  where $T$ is the temperature and $r$ is the radial distance \citep{mihalas_1984_aa}.

The radiative opacity depends on the energy of the photons and three other quantities shared with the stellar equation of state -- the temperature $T$, density $\rho$, and composition vector. As a result, $\kappa$ and its partial derivatives with respect to thermodynamic quantities can impact a wide range of stellar phenomena. 

For example, the $\kappa$-mechanism drives changes in the luminosity of many types of pulsating variable stars \citep{eddington_1926_aa,cox_1980_aa,aerts_2010_aa}. In regions where the opacity increases with temperature (e.g., where hydrogen and helium are partly ionized), the atmosphere becomes unstable against pulsations \citep{hansen_2004_aa,kippenhahn_2012_aa, das_2020, kurtz_2022_aa}. Hydrogen ionization drives the pulsations of Mira variables \citep{fox_1982_aa,fadeyev_2022_aa} and red supergiants \citep{heger_1997_aa,yoon_2010_aa}, the high-overtone, low-degree, non-radial pressure modes of rapidly oscillating Ap stars \citep{kurtz_1982_aa,shibahashi_1993_aa,holdsworth_2021_aa,bigot_2002_aa}, and ZZ Ceti variables \citep{landolt_1968_aa,corsico_2019_aa}. Helium ionization drives pulsations in RR Lyrae variables \citep{smith_2004_aa,ngeow_2022_aa} and $\delta$~Scuti variables \citep{balona_2018_aa,bowman_2018_aa,guzik_2021_aa,murphy_2023_aa}, DBV white dwarf variables 
\citep{corsico_2019_aa,saumon_2022_aa},
and is furthermore responsible for acoustic glitches in solar-like oscillators \citep{gough_1990, basu_2004, mazumdar_2014, verma_2017, verma_2019, saunders_2023}. Other opacity increases from 
the iron group elements (Cr, Fe, Ni and Cu)  at temperatures of $\simeq$\,2$\times$10$^5$\,K and densities of $\simeq$10$^{-7}$\,g\,cm$^{-3}$ are likely the cause of pulsations in B-type stars \citep{townsend_2005_aa,aerts_2010_aa, guzik_2018_aa, shi_2023_aa}, and $\beta$-Cephei variables (e.g., $\beta$ Centauri, $\beta$ Centauri, $\gamma$ Pegasi and $\nu$ Eridani)
where $\kappa$ may account for differences between the observed and calculated pulsation periods 
\citep{daszynska-daszkiewicz_2009_aa,daszynska-daszkiewicz_2010_aa,cugier_2012_aa,walczak_2015_aa}.

Another example is the solar structure \citep{basu_2016_aa,buldgen_2019_aa,christensen-dalsgaard_2021_aa} where the ionization of C, N, O, Ne, and Fe group elements near the base of the solar convection zone induces bound-free transitions that can be a source of the discrepancy between inferences from helioseismology measurements and solar photosphere composition determinations \citep{turck-chieze_2004_aa,bahcall_2006_aa,guzik_2008_aa,basu_antia_2008_aa,guzik_mussack_2010_aa,neuforge-verheecke_2001_aa,neuforge-verheecke_2001_ab,krief_2016_aa,pradhan_2023_ab}.

Differences between theoretical opacities and observation-adjusted opacities are discussed in \cite{eddington_1926_aa,daszynska-daszkiewicz_2017_aa,guzik_2018_aa,daszynska-daszkiewicz_2023_aa}. Considerations of negative hydrogen ion transitions \citep{wildt_1939_aa,chandrasekhar_1944_aa,chandrasekhar_1946_aa,ohmura_1960_aa,ohmura_1964_aa,doughty_1966_aa}, bound-bound transitions \citep{mayer_1947_aa, meyerott_1951_aa, moszkowski_1951_aa}, bound-free transitions (photoionization), free-free transitions (inverse bremsstrahlung) and electron scattering \citep{vitense_1951_aa,schwarzschild_1958_aa,vardya_1964_aa,jin_1982_aa,meyer-hofmeister_1982_aa} have been followed by an extensive literature reporting atomic radiative opacity calculations. 

Examples include 
Los Alamos/OPLIB 
\citep{cox_1962_aa,cox_1965_aa,cox_1970_aa,cox_1970_ab,cox_1976_aa,hubner_1977_aa,weiss_1990_aa,magee_1995_aa,colgan_2016_aa},
the Opacity Project
\citep{seaton_1987_aa,seaton_1994_aa,seaton_2004_aa,seaton_2005_aa,badnell_2005_aa, pradhan_2023_aa,nahar_2023_aa,pradhan_2023_ab,zhao_2023_aa},
Livermore OPAL
\citep{rogers_1992_aa,iglesias_1993_aa,iglesias_1996_aa},
analytic fitting
\citep{christy_1966_aa,iben_1975_aa,stellingwerf_1975_aa,stellingwerf_1975_ab}, 
the Scotland model
\citep{carson_1968_aa,carson_1976_aa},
the OPAS model 
\citep{blancard_2012_aa,le-pennec_2015_aa,mondet_2015_aa},
the hybrid model SCO-RCG
\citep{iglesias_2012_aa,pain_2015_aa,pain_2017_aa,pain_2019_aa},
for actinides \citep{fontes_2023_aa,flors_2023_aa},
and for He-dominated compositions, 
He$^-$ free–free \citep{somerville_1965_aa,john_1994_aa}, 
He$^{+}_{2}$ bound–free and free-free \citep{ignjatovic_2009_aa}, 
He Rayleigh scattering \citep{iglesias_2002_aa,rohrmann_2018_aa}, and 
triple-He collisions \citep{kowalski_2014_aa,blouin_2019_aa,saumon_2022_aa}.

In this article we add a novel contribution to this canon by providing open access to an expanded set of 1194 Type-1 Rosseland-mean opacity tables for four different heavy element mixtures, with improvements to the composition range, table coverage, and table resolution. Section~\ref{s.atomic_tables} describes the methods used to construct the opacity tables and highlights a few representative results. Section~\ref{s.models} implements the new opacity tables in \MESA\ and applies them to the helioseimology and neutrino fluxes of standard solar models, and Section~\ref{s.conclusion} offers concluding remarks. Appendices A to C detail the implementation and verification of the new Los Alamos opacity tables in \MESA.

Important symbols are defined in Table \ref{table:list-of-symbols}.
In this article ``log" refers to the base-10 logarithm; where the natural logarithm is intended, we use “ln”.

\startlongtable
\begin{deluxetable}{clc}
  \tablecolumns{3}
  \tablewidth{1.0\apjcolwidth}
  \tablecaption{Important symbols.
   \label{table:list-of-symbols}}
  \tablehead{\colhead{Name} & \colhead{Description} & \colhead{Appears}}
  \startdata
$\alpha$        & $^4$He particle                      & \ref{s.helio}  \\
$a$             & Radiation constant = 4$\sigma/c$     & \ref{s.atomic_tables}  \\
$A$             & Atomic number                        & \ref{s.intro}  \\
$c$             & Speed of light in the medium         & \ref{s.atomic_tables}  \\
$c_{s}$         & Acoustic sound speed                & \ref{s.helio}  \\
$G$             & Gravitational constant               & \ref{s.atomic_tables}  \\
$E$             & Energy                               & \ref{s.atomic_tables}  \\
$\gamma$        & Photon                               & \ref{s.helio}  \\
$\Gamma_3$      & $\equiv d{\rm ln}T/d{\rm ln}\rho|_S+1$ & \ref{s.atomic_tables}  \\
$h$             & Planck constant                      & \ref{s.atomic_tables}  \\
$k_{\rm B}$     & Boltzmann constant                   & \ref{s.atomic_tables}  \\
$\kappa$        & Opacity                              & \ref{s.intro}  \\
$\kappa_{\nu}$  & Monochromatic opacity                & \ref{s.atomic_tables}  \\
$\kappa_{P}$    & Planck mean opacity                  & \ref{s.atomic_tables}  \\
$\kappa_{R}$    & Rosseland mean opacity               & \ref{s.atomic_tables}  \\
$\lambda$       & Mean free path                       & \ref{s.atomic_tables}  \\
$L$             & Luminosity                           & \ref{s.atomic_tables}  \\
$M$             & Stellar mass                         & \ref{s.atomic_tables}  \\
$\nu$           & Frequency                            & \ref{s.atomic_tables}  \\
$\nu_{e}$       & Electron neutrino                    & \ref{s.helio}  \\
$\nabla_{\text{rad}}$ & Radiative temperature gradient & \ref{s.atomic_tables}  \\
$r$             & Radial coordinate                    & \ref{s.intro}  \\
$R$             & Opacity coordinate                   & \ref{s.atomic_tables}  \\
$R_o$           & Stellar radius                       & \ref{s.atomic_tables}  \\
$\rho$          & Mass density                         & \ref{s.intro}  \\
$\Phi$          & Neutrino Flux                        & \ref{s.helio} \\
$P$             & Pressure                             & \ref{s.atomic_tables}  \\
$\sigma$        & Stefan–Boltzmann constant            & \ref{s.atomic_tables}  \\
$S$             & Entropy per gram                     & \ref{s.atomic_tables}  \\
$t$             & Time                                 & \ref{s.atomic_tables}  \\
$T$             & Temperature                          & \ref{s.intro}  \\
$X$               & Hydrogen mass fraction               & \ref{s.atomic_tables}  \\
$Y$               & Helium mass fraction                 & \ref{s.atomic_tables}  \\
$Z$               & Metal mass fraction                  & \ref{s.atomic_tables}  \\
Z             & Atomic charge                        & \ref{s.atomic_tables}  \\
  \enddata
\tablenotetext{}
{\hsize \apjcolwidth {\bf Note:} 
Some symbols may be further subscripted, for example,
by an $\mathrm{a}$ (indicating an absorption quantity),
by an $\mathrm{s}$ (indicating a scattering quantity), or
by a $\mathrm{t}$ (indicating a total quantity).
}
\end{deluxetable}


\section{Stellar Opacity}\label{s.atomic_tables}

The total monochromatic radiative opacity $\kappa_{\nu, {\rm t}}$ is the sum  of the absorption opacity $\kappa_{\nu, {\rm a}}$ and scattering opacity $\kappa_{\nu, {\rm s}}$ at a specific $T$, $\rho$ and composition. The mean (or gray) opacity represents, in a single number, the tendency of a material to absorb and scatter radiation of all frequencies. Two common mean opacities are the Planck mean (or emission mean) and Rosseland mean opacities. Other examples include a flux-weighted mean and an absorption mean. These various means arise in order to obtain correct values for a particular frequency-integrated physical quantity, such as the radiation flux or energy \citep{cox_1968_aa,mihalas_1978_aa,cowan_1981_aa,mihalas_1984_aa,hansen_2004_aa,kippenhahn_2012_aa,huebner_2014_aa,fontes_2015_aa,fontes_2023_aa}.

The Planck mean opacity $\kappa_P$ yields the correct value for the integrated thermal emission for an optically thin plasma
\begin{equation}
\kappa_P = \frac{\int_0^\infty \ \kappa_{\nu, {\rm a}} \ B_\nu(T) \ d\nu}{\int_0^\infty B_\nu(T) \ d\nu}
\ ,
\end{equation}
where $B_\nu(T)$\,=\,$2 h \nu^3$/$c^2$\,$\cdot$\,1/($\exp(h\nu /k_{\rm B}T) - 1$) is the Planck function, $k_{\rm B}$ is the Boltzmann constant, $h$ is the Planck constant, $c$ is the medium's speed of light, and $\nu$ is the photon frequency. Note $\kappa_P$ is calculated only from $\kappa_{\nu, {\rm a}}$. This weighting function peaks at $h\nu$\,$\simeq$\,2.8\,$k_{\rm B}T$, indicating where $\kappa_{\nu, {\rm a}}$ is most strongly sampled.

The Rosseland mean opacity $\kappa_R$ yields the correct value for the integrated energy flux of an optically thick plasma
\begin{equation}
\frac{1}{\kappa_R} = \frac{\int_0^\infty \frac{1}{\kappa_{\nu,t}} \ \frac{dB_\nu(T)}{dT} \ d\nu}{\int_0^\infty \frac{dB_\nu(T)}{dT} \ d\nu} 
\ .
\end{equation}
The use of a harmonic average means the individual contributions (bound-bound, bound-free, free-free, scattering) cannot be averaged first and then added to obtain the proper mean value. This weighting function peaks at $h\nu$ $\simeq$\,3.8 $k_{\rm B}T$, indicating where $\kappa_{\nu, {\rm t}}$ is strongly sampled.

The radiation transport equation in the gray-diffusion approximation \citep[e.g.,][]{mihalas_1978_aa}
\begin{equation}
\frac{\partial E_{r}}{\partial t} = \grad \cdot \left ( \frac{c/3}{\rho \kappa_R} \grad E_{r} \right ) + c\rho \kappa_P (a T_e^4 - E_{r} ) 
\end{equation}
admits two mean free paths;  $\lambda_{R}$\,=\,1/$\rho$$\kappa_R$ (used in the diffusion coefficient term) and $\lambda_{P}$\,=\,1/$\rho$$\kappa_P$ (used in the radiation-electron coupling term). The two mean free paths can differ by several orders of magnitude due to the different averaging prescriptions. Here $E_{r}$ is the radiation energy, $a$ is the radiation constant, and $T_e$ is the electron temperature. 
In stellar evolution models, $\kappa_{R}$ is used for the radiative temperature gradient 
\begin{equation} \label{eq1}
\nabla_{\text{rad}} = \left( \frac{d \ln T}{d \ln P} \right)_{\text{rad}} = \frac{3 \kappa_R L P}{16 \pi a c G M_r T^4}
\ ,
\end{equation}
where $L$ is the luminosity, $P$ is the pressure, $G$ is the gravitational constant, and $M_r$ is the mass contained within radius $r$. $\nabla_{\text{rad}}$ in the mixing-length theory of convection (MLT) determines convective stability, the radiative flux in convective regions, and the actual temperature gradient in radiative regions. 

Radiative opacities depend on the composition, which can change rapidly in mass or time due to nuclear reactions, diffusion or chemical mixing. A typical approach is to adopt a fixed solar abundance mixture of metals with mass fraction $Z$, hydrogen mass fraction $X$, and helium mass fraction $Y$. Individual Rosseland mean opacity tables are calculated for various combinations of $X$ and $Z$ where $Y$\,=\,1 $-$ $X$ $-$ $Z$, and the chemical elements that compose $Z$ have a fixed distribution. Tabulated Rosseland mean opacity values computed with a fixed metal distribution are an acceptable alternative to calculating self-consistent Rosseland mean opacities that reflect the exact metal distribution of a stellar model as it evolves. The adoption of self-consistent Rosseland mean opacity tables results in a $\lesssim$\,2\% change to the total opacity in a stellar model \citep{hui-bon-hoa_2021_aa}.

We use \MESA's \texttt{kap} module to implement the new atomic radiative opacity tables and compare them to previous atomic opacity releases. The \texttt{kap} module builds opacity tables by combining the radiative opacities and electron conduction opacities 
\begin{equation}
\frac{1}{\kappa} = \frac{1}{\kappa_{R}} + \frac{1}{\kappa_{\rm cond}}
\end{equation}
where $\kappa_{\rm cond}$\,=\,16$\sigma$T$^{3}$/$\rho$$K$  
converts an electron conductivity $K$ to an opacity.
Electron conduction opacities are tabulated for atomic charges 1\,$\leq$\,Z\,$\leq$\,60 between 3\,$\leq$\,log\,$T$\,$\leq$\,10  and -6\,$\leq$\,log\,$\rho$\,$\leq$\,11.5 \citep{cassisi_2007_aa,blouin_2020_aa}.

The cores of stellar models mostly evolve along lines of constant specific radiation entropy $S_{\rm rad}$\,$\sim$\,$T_{\rm c}^3$/$\rho_{\rm c}$\,$\sim$\,$M$$^2$ for a fixed non-degenerate stellar mass $M$.
Atomic and molecular radiative opacities in \MESA\ are tabulated using the conventional OPAL log\,$R$\,--\,log\,$T$ format, where $R$\,=\,$\rho$/$T_6^3$\ which scales as $\sim$1/$S_{\rm rad}$. The advantage of using $T_6$\,=\,$T$/10$^6$ K and $R$ is that the range of interest for stellar physics can be covered by a rectangular array in these variables \citep{bahcall_1988_aa}.

\begin{figure*}[!htb]
\centering
\includegraphics[width=7.0in]{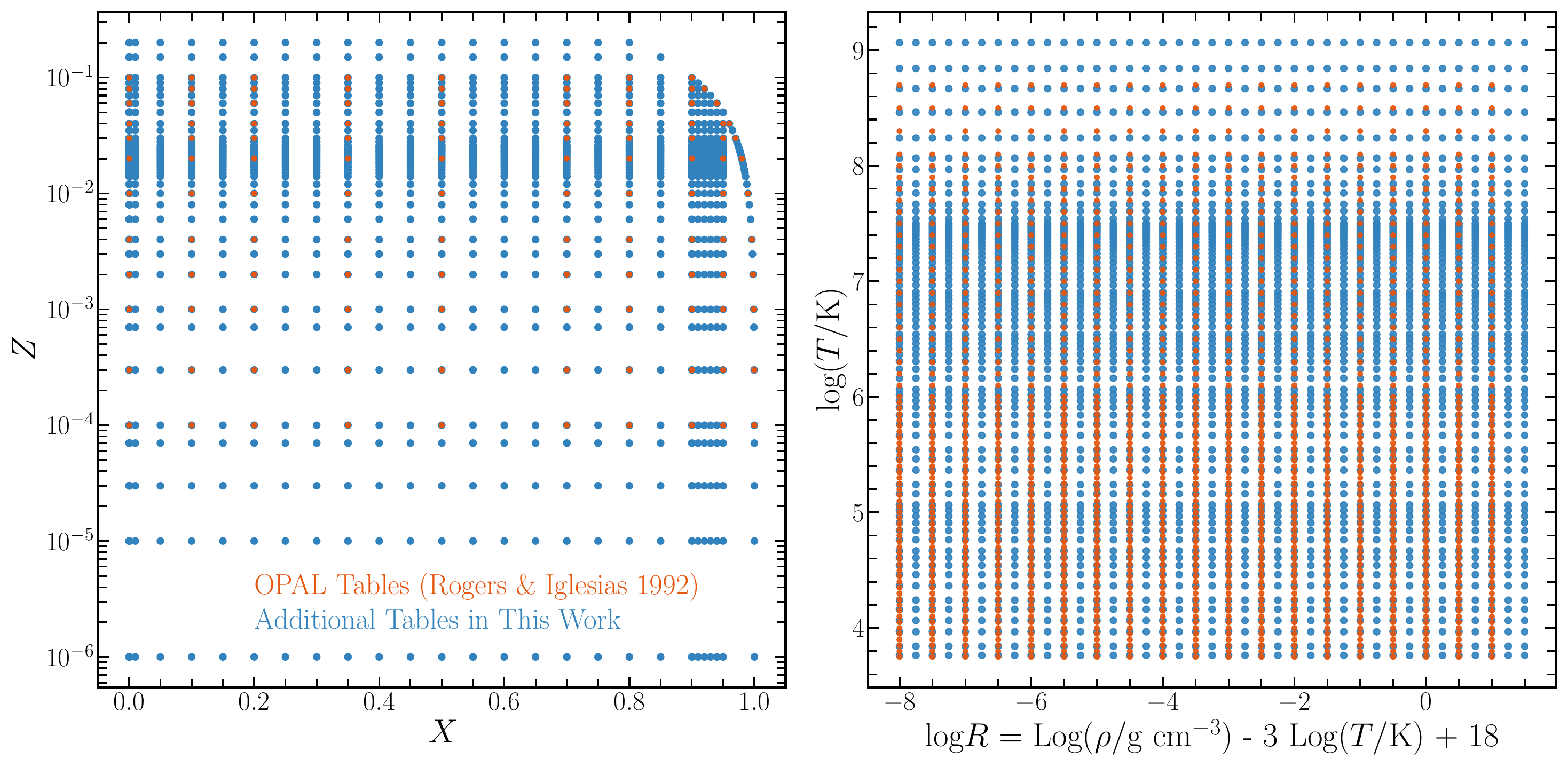} 
\caption{Location of each Type-1 opacity table in the $X$--$Z$ plane (left panel) and the $\log T$ --$\log R$ plane (right panel). Orange circles mark the location of the 126 OPAL Type-1 tables \citep{rogers_1992_aa}. Blue circles mark the location of the new 1194 Type-1 opacity tables. }
\label{f.fig1}
\end{figure*}

Molecular transitions (e.g., H$_2$, H$_2$O, TiO, CO) are the primary source of the radiative opacity for log\,$T$\,$\leq$\,4.5. In this regime the \texttt{kap} module provides radiative opacities from \citet{ferguson_2005_aa}. The \texttt{kap} module also supplies a privately communicated set of \citet{freedman_2008_aa} low-temperature opacities,  and supports \AE SOPUS opacities \citep{marigo_2009_aa} for arbitrary chemical mixtures \citep[see][for details]{jermyn_2023_aa}.
\MESA\ currently defaults to blending the higher temperature and lower temperature opacity tables between 3.80\,$\le$\,$\log T$\,$\le$\,3.88 \citep[see][]{paxton_2013_aa}.
Compton scattering dominates the radiative opacities for log\,$T$\,$\geq$\,8.7, where the \texttt{kap} module uses \citet{poutanen_2017_aa}.
For intermediate temperatures, atomic radiative opacities have been supplied by OPAL \citep{iglesias_1993_aa,iglesias_1996_aa} and OP \citep{seaton_2005_aa,badnell_2005_aa}.  Type-1 OPAL tables, the subject of this article, have a fixed heavy element metal distribution (e.g., H burning). Type-2 OPAL tables provide radiative opacities for C and O rich mixtures \citep[e.g., He burning,][]{iglesias_1993_aa}.

\subsection{OPLIB Database}\label{s.atomic_code}

The Los Alamos OPLIB opacity database has been publicly available for more than forty years, and is currently accessible at the website \url{http://aphysics2.lanl.gov/opacity/lanl}. The database contains monochromatic opacities for the first thirty elements of the periodic table, using a photon-energy grid of 14,900 points in the variable $u$\,=\,$h\nu/k_{\rm B}T$, with values ranging from 10$^{-4}$\,$\le$\,$u$\,$\le$\,30,000. The website can produce monochromatic, multigroup and gray opacities for either pure elements or arbitrary mixtures. The most recent database release \citep{colgan_2016_aa}, which is referred to simply as ``OPLIB" in the comparisons provided in this article, was generated with the \texttt{ATOMIC} code, while the previous release, which is referred to as ``OPLIB-L", was generated with the \texttt{LEDCOP} code \citep{magee_1995_aa}.

Since this article focuses on the latest OPLIB release, which is intended to supersede the OPLIB-L release, we provide a brief summary of the relevant computational methods. \texttt{ATOMIC} is a  multi-purpose plasma modeling code \citep{magee_2004_aa,hakel_2006,fontes_2015_aa} that can be run in local thermodynamic equilibrium (LTE) or non-LTE mode to calculate the atomic-state populations. In this article, we focus on LTE populations. These populations are combined with atomic data, e.g. oscillator strengths and photoionization cross-sections, to obtain the monochromatic opacities, which are constructed from the standard four contributions: bound–bound, bound–free, free–free, and scattering.

When calculating the populations, \texttt{ATOMIC} uses the ChemEOS model to account for the effects of the plasma environment on the equation of state \citep{hakel_2004_aa,hakel_2006,kilcrease_2015_aa}.
Whereas OPAL is based on the physical picture that treats plasma as a collection of nuclei and electrons Coulomb-interacting in the grand canonical ensemble, ChemEOS uses the chemical picture in which the free electrons and the various ion stages are identified as individual species in the canonical ensemble; the associated Helmholtz free energy is then minimized in ChemEOS to yield the species’ populations. From that perspective ChemEOS is similar to OP, with which it also shares the use of occupation probabilities to model the gradual dissolution of ion-stages’ bound states by the plasma environment. ChemEOS differs from OP in the precise form of the occupation probabilities due to its adopted plasma microfield model, and in the details of excluded-volume considerations for the ion-stages’ finite sizes. Furthermore, in order to extend the validity of the chemical picture to high densities ChemEOS adopts a model for the species’ Coulomb interactions that smoothly bridges the transition from the Debye-Hückel limit at low densities to the Thomas-Fermi limit in the strongly-coupled high-density regime.

For the latest OPLIB release, \texttt{ATOMIC} used semi-relativistic Hartree-Fock atomic data that were generated with the Los Alamos suite of atomic physics codes \citep{fontes_2015_aa}. Improvements over the previous (\texttt{LEDCOP}) OPLIB release include the addition of 24 isotherms to reduce interpolation errors, extending the calculations to higher densities and the inclusion of significantly more lines via a histogram method \citep{abdallah_2007}. Additional details about the latest OPLIB release, as well as comparisons with other opacity databases, are provided in \cite{colgan_2016_aa}. These improvements have made possible the generation of higher fidelity opacity tables for pre-tabulated ($X$,$Y$,$Z$) mixtures of interest, as compared to the pre-tabulated OPLIB-L tables publicly available at  \url{http://aphysics2.lanl.gov/opacity/lanl}. We detail these improvements in the next section.

Before leaving this section, we provide some general information about uncertainties in the calculated opacities. Uncertainties in Rosseland mean opacity are caused by uncertainties in the calculation of the fundamental atomic cross sections, plasma effects caused by perturbing ions, and computational limitations. For example, measurements of fundamental cross sections are usually carried out on neutral atoms, rather than on charged ions, due to the difficulty in preparing a sample in a specific ion stage and because of the myriad possibilities of excited levels. On the other hand, cross sections of neutral atoms are more difficult to calculate accurately because of the many-body, electron-electron interactions. Thus, comparison of calculations with measured cross sections for neutral species can provide an upper bound on cross section uncertainties. For Sun-like conditions, estimates of the uncertainty in the opacity are $\simeq$5\% when electron scattering dominates at high $T$ and low $\rho$  \citep{huebner_2014_aa}. As the $\rho$ increases and free-free processes become more important, the uncertainty is less than $\simeq$10\%. As $T$ decreases and bound-free processes become important, the uncertainty increases to $\simeq$20\%. As $T$ decreases further, bound-bound processes can contribute, and the uncertainty rises to $\simeq$30\% \citep{huebner_2014_aa}. From atomic-theory considerations, the uncertainties in the calculated cross sections, particularly those that involve bound electrons, progressively decrease as the ionic charge increases toward the limiting case of one-electron (hydrogenic) ions, provided that plasma effects do not become too important. Such conditions exist for a variety of astrophysical applications, such as the solar modeling discussed in Section~\ref{s.models}.

\subsection{New OPLIB Opacity Tables}\label{s.tables}

Figure~\ref{f.fig1} shows the OPLIB tables in the $X$--$Z$ plane (left panel). Previous opacity releases by OP and OPAL contain 126 tables spanning $Z$\,=\,10$^{-4}$ to $Z$\,=\,0.1, and $Z$\,=\,0.0. We have improved on these grids in three ways. First, we expand the range to $Z$\,=\,10$^{-6}$ to $Z$\,=\,0.2, allowing for improved accuracy of opacities at low and high $Z$. Second, we substantially increase the table density in the range  $Z$\,=\,10$^{-4}$ to $Z$\,=\,0.1 enhancing the accuracy of opacities drawn from interpolations across neighboring metallicities.  Third, we add opacity tables between $X$\,=\,0 and $X$\,=\,0.1. The addition of tables at $X$~$= 10^{-2}, 10^{-3}, 10^{-4}, 10^{-5}, 10^{-6}$ can, for example, improve stellar models with thin H-depleted mixtures, such as in the atmospheres of hot subdwarfs or other stripped stars. These new OPLIB tables encompass 1194 individual tables, an order of magnitude larger in number than any previous opacity table release, and span regimes where previous opacity tables have not existed.

Figure~\ref{f.fig1} also shows the OPLIB tables in the $\log T$--$\log R$ plane (right panel). OP and OPAL radiative opacity tables are tabulated for 3.75~$\leq$~log($T$/K)~$\leq$~8.7, and $-$8~$\leq$ ~log(R)~$\leq$~1, with 70 tabulated points in log($T$/K) and 19 tabulated points in log($R$). The new OPLIB radiative opacity tables are tabulated over a larger space of 3.764~$\leq$~log($T$/K)~$\leq$ 9.065, and $-$8~$\leq$~log($R$)~$\leq$~1.5 with 74 tabulated points in log($T$/K) and 39 tabulated points in log($R$).  Table values are written to four decimal place precision, an improvement over other works which write values to three decimal place precision. Each opacity table is calculated with a mixture of 25 elements: H, He C, N, O, F, Ne, Na, Mg, Al, Si, P, S, Cl, Ar, K, Ca, Sc, Ti, V, Cr, Mn, Fe, Co, Ni.

\subsection{OPLIB Opacity Tables in \MESA}\label{s.inmesa}

The 126 OPAL and OP opacity tables are pre-processed via smoothing and spline-fitting routines provided in \citet{seaton_1993_aa} to ensure smooth opacity derivatives. When run through these routines, OP and OPAL tables are smoothed and interpolated from 70 log($T$/K) $\times$ 19 log($R$) point tables spanning 3.75 $\leq$ log($T$/K) $\leq$ 8.70 and $-$8.0 $\leq$ log($R$) $\leq$ 1.0 into evenly spaced 138 log($T$/K) $\times$ 37 log($R$) opacity tables using bi-cubic splines. We take a similar approach by applying bi-cubic splines without a smoothing filter to the OPLIB tables, which interpolate our 74 log($T$/K) points up to $213$ points. Further discussion of the interpolation method along with comparisons between the raw and interpolated tables are provided in Appendix~\ref{appendix.A}.

\MESA's \code{kap} module computes the radiative opacity given $\rho$, $T$, $X$, $Z$ from a cell. For a fixed composition opacity table in the $X$--$Z$ plane, \MESA\ interpolates between $\rho$ and $T$ values with an on-the-fly bi-cubic spline. The spline returns $\kappa_{R}(\rho,T)$, and its partial derivatives, $\partial{\kappa_{R}(\rho,T)}/\partial{T}$ and $\partial{\kappa_{R}(\rho,T)}/\partial{\rho}$.

\MESA\ currently offers two choices for interpolating between opacity tables in $X$--$Z$: linear or monotonic Hermite cubic spline functions \citep[see][]{paxton_2011_aa}. The default is linear interpolation, as exemplified in \MESA's test suite \citep{wolf_2023_aa}. In this article we activate cubic interpolation in \MESA's inlist controls to return $\kappa_{R}(\rho,T,X,Z)$: 
\begin{itemize}\tightitems
\item[] \code{cubic\_interpolation\_in\_X = .true.} 
\item[] \code{cubic\_interpolation\_in\_Z = .true.}
\end{itemize}
 
Appendix~\ref{appendix.B} explores the impact of adopting cubic versus linear interpolation of opacity data tables across $X$--$Z$ in \MESA. We find that linear interpolation systematically under predicts the opacity as compared to cubic interpolation, and recommend cubic interpolation be activated when using \MESA. We leave further exploration and improvements to \MESA's opacity interpolation methods to future work.

\begin{figure*}[!htb]
\centering
\includegraphics[width=6.5in]{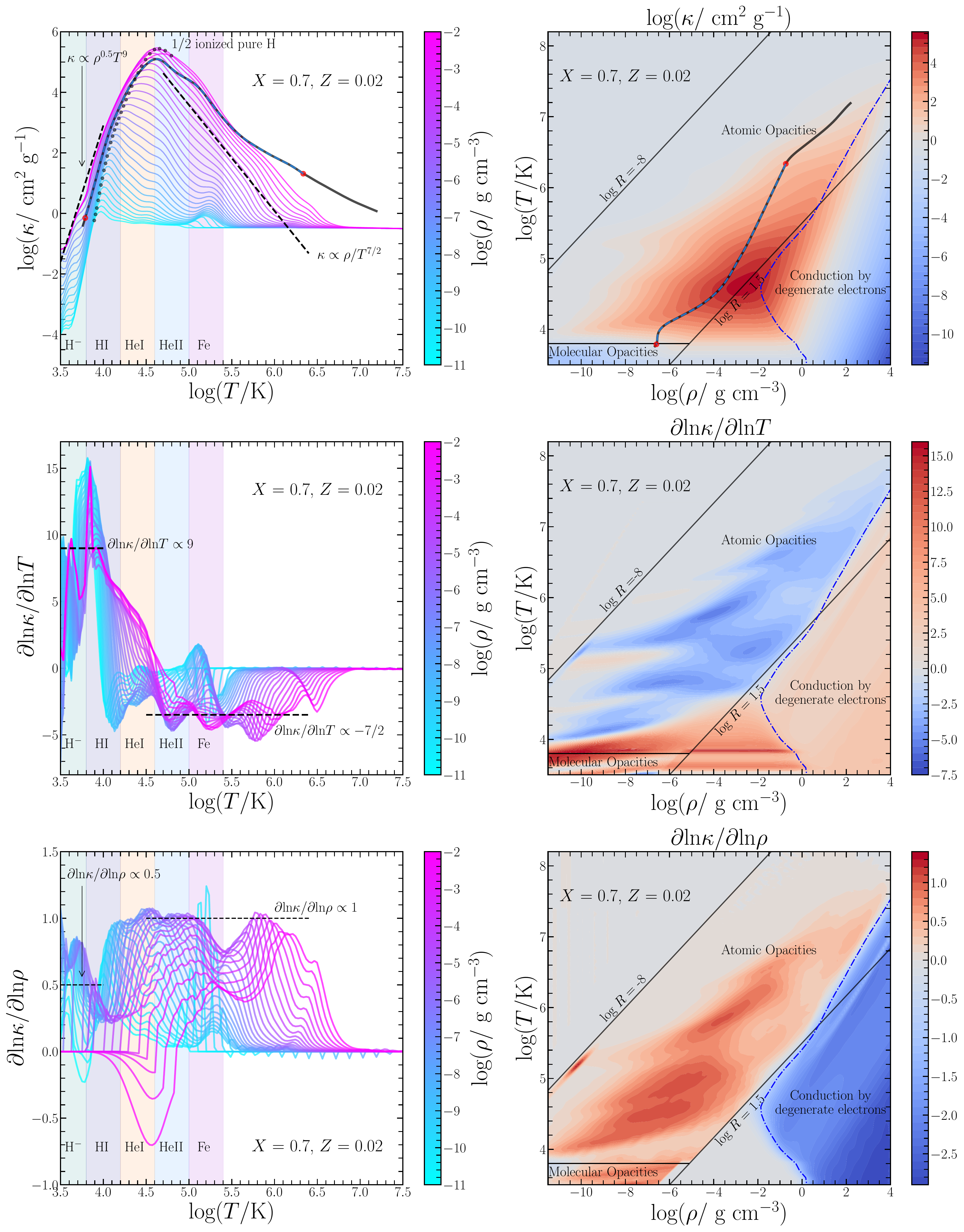} \\
\caption{Opacities and partial derivatives with respect to temperature and density for a $X$\,=\,0.7, $Z$\,=\,0.02 \citet{grevesse_1998_aa} abundances generated from OPLIB radiative opacities and \MESA's \texttt{kap} module. {\it Left column}: These quantities as a function of $T$ for different $\rho$ (colorbar). Dashed black lines show scaling relations for H$^-$ and Kramers opacities. Black dots mark the locations where a pure H composition is 1/2 ionized. Colored regions show thermal ionization stages of key elements for nondegenerate material. {\it Right column}:  These quantities as contours in the $\rho-T$ plane.  Regions where different opacity sources dominate are labeled, as are the log\,R\,=\,$-$8 and log\,R\,=\,1.5 table limits of the OPLIB radiative opacities. {\it Both columns}: Black curves show the profile of a standard solar model (see Section~\ref{s.models}). Red circles on the solar profile mark the inner and outer boundaries of the convective region and are connected with a dashed blue curve.
}
\label{f.panelA}
\end{figure*}

Figure~\ref{f.panelA} shows $\kappa$ and its partial derivatives $\partf{\kappa}{T}$ and $\partf{\kappa}{\rho}$ generated by the OPLIB radiative opacities and \MESA's \texttt{kap} module for a $X$\,=\,0.7, $Z$\,=\,0.02 \citet{grevesse_1998_aa} abundance mixture. The left column shows these quantities as a function of $T$ for different $\rho$. The dependence of $\kappa$ with $T$ in the upper left plot can be approximated with three temperature ranges. At low temperatures, log($T$/K)\,$\lesssim$\,4, the H$^{-}$ opacity dominates and scales as $\kappa$\,$\sim$\,$\rho^{1/2} T^9$ \citep{hansen_2004_aa} as shown by the labeled dotted black line in the upper left plot of Figure~\ref{f.panelA}. At higher temperatures, 4\,$\lesssim$\,log($T$/K)\,$\lesssim$\,8, the free-free (inverse thermal bremsstrahlung) and bound-free (radiative recombination) opacity dominates and scales as $\kappa$\,$\sim$\,$\rho T^{7/2}$ \citep{kramers_1923_aa,gaunt_1930_aa} also shown by a labeled dotted black line. At still higher temperatures, log($T$/K)\,$\gtrsim$\,8, all the opacity curves converge to $\kappa \sim$\,constant, the flat plateau at the foot of the ``$\kappa$ mountain" \citep{kippenhahn_1990_aa}. When the atoms are completely ionized, then there are only two sources of opacity, free-free and Compton scattering. For low densities, Compton scattering dominates $\kappa_R$ and does not impact $\kappa_P$. For high densities, free-free transitions dominate $\kappa_R$ and $\kappa_P$. 

Two additional features of the $\kappa$ curves in the upper left plot of Figure~\ref{f.panelA} are noteworthy. The first is that the location of peak $\kappa$ shifts towards higher $T$ as $\rho$ increases. The H$^-$ opacity depends on the abundance of neutral hydrogen. A pure atomic H composition is 1/2 ionized by the Saha equation when \citep{hansen_2004_aa}
\begin{equation} \label{eq.half_h}
\rho = 8.02\times10^{-9} \ T^{3/2} \ \exp(-1.578\times10^5 / T)
\ .
\end{equation}
This locus of points is shown by the black circles and reflects the broad trend in the location of peak $\kappa$, where the dominant opacity smoothly transitions from H$^-$ to bound-free. The second feature is the Fe group opacity bump (or $Z$ bump) at log($T$/K)\,$\simeq$\,5.35, centered in the colored region denoting the thermal ionization regime of Fe, most prominent at low densities (blue curves).

The standard solar model shown in upper left plot of Figure~\ref{f.panelA}, which is detailed in Section~\ref{s.models}, begins at the photosphere on the left with log($T$/K)\,=\,3.762 and log($\rho$/g cm$^{-3}$)\,=\,$-$6.66. Progressing inwards, towards higher $T$, the opacity sharply increases due to ionization of elements with low-lying first ionization stages (e.g., Na, Mg, Al, K, Ca, and H). These stages provide electrons for H$^{-}$ formation, and $\kappa$ rises by several orders of magnitude until it reaches a maximum at approximately the half-ionization curve for pure H, when most H is ionized and thus is not available for H$^{-}$ formation \citep{kippenhahn_1990_aa,hansen_2004_aa}. Progressing further inwards, bound-free transitions become the main source of opacity and still further inwards free-free transitions dominate. The solar core remains in the domain of free-free transitions \citep{kippenhahn_1990_aa}. 

The upper right plot of Figure~\ref{f.panelA} shows contours of $\kappa$ in the $\rho - T$ plane over a larger range of $T$ and $\rho$ than in the corresponding upper left plot. The red $\kappa$~mountain is prominent. The overlaid standard solar model profile shows that high-energy photons generated in the solar core follow one trajectory in traversing a face of the $\kappa$ mountain before eventually being released as lower-energy photons at the photosphere. Another face of the $\kappa$ mountain, towards larger $\rho$, is bounded by the region where $\kappa_{\rm cond}$ dominates the opacity. Note the log($R$)\,=1.5 opacity tables define the peak of the $\kappa$~mountain.

The partial derivatives of the opacity with respect to temperature $\partf{\kappa}{T}$ and with respect to density $\partf{\kappa}{\rho}$ are useful for assessing excitation of a $\kappa$-mechanism \citep{saio_1993_aa}
\begin{equation}
\frac{{\rm d}}{{\rm d}r} \left ( \frac{\partial \kappa}{\partial T} + \frac{\partial\kappa}{\partial\rho}  \frac{1}{\Gamma_3 - 1} \right ) > 0
\ ,
\end{equation}
where $\Gamma_3$\,$\equiv$\,$d{\rm ln}T/d{\rm ln}\rho|_S+1$ is the third dimensionless adiabatic exponent from the equation of state. Opacity derivatives are also useful for constructing the Jacobian matrix of derivatives for Newton-like iterations to solve the stellar evolution hydrodynamic equations.

\begin{figure*}[!htb]
\centering
\includegraphics[width=6.5in]{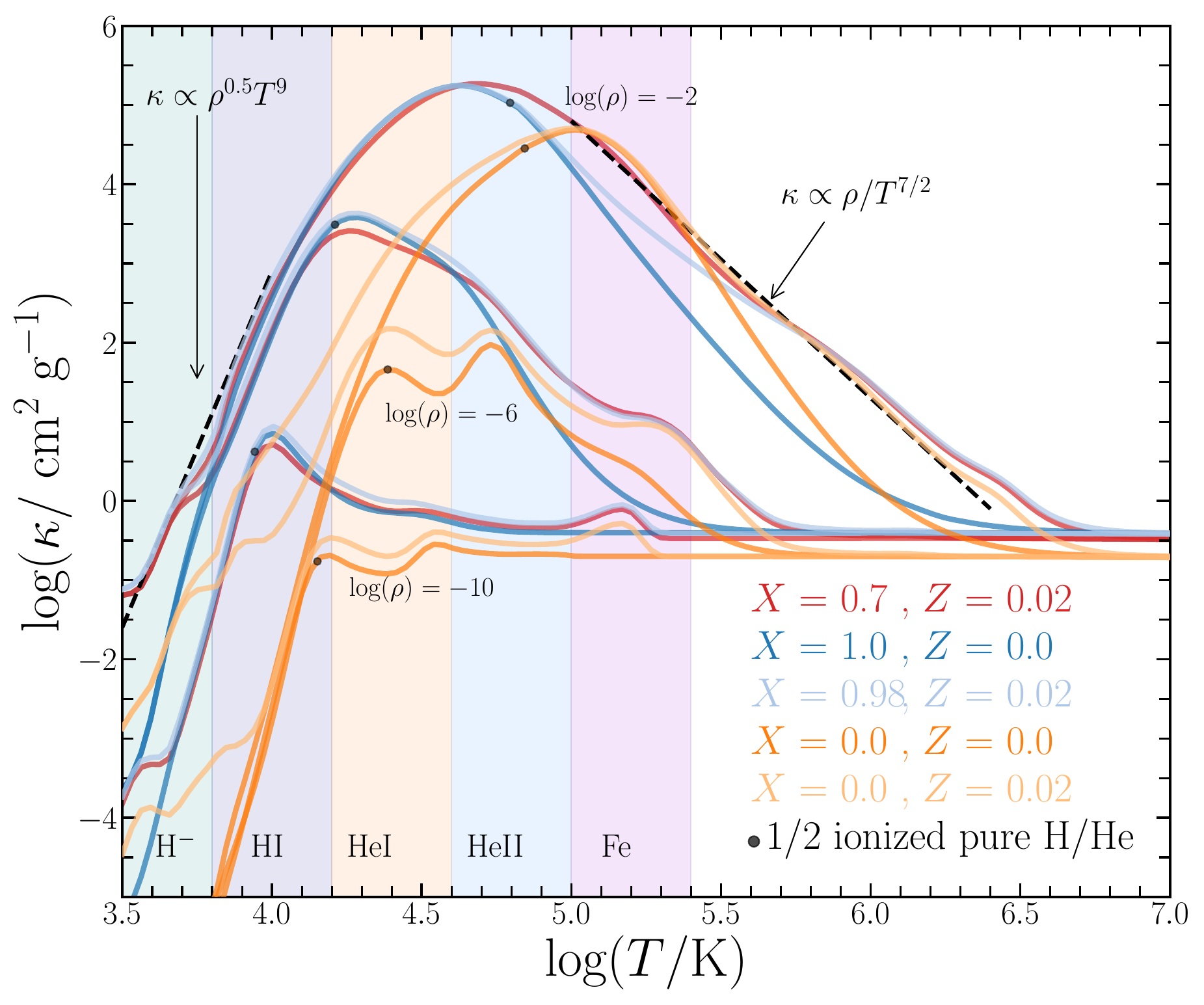} \\
\caption{Opacity as a function of $T$ for three values of $\log(\rho)$ (black labels) and five compositions  with \citet{grevesse_1998_aa} abundances (colored curves), generated from OPLIB radiative opacities and \MESA's \texttt{kap} module. Dashed black lines show scaling relations for H$^-$ and Kramers opacities. Black circles mark the half-ionization points for the pure atomic H composition (dark blue curves, Equation~\ref{eq.half_h}) and the pure atomic He composition (dark orange curves, Equation~\ref{eq.half_he}). Colored regions show thermal ionization stages of key elements for nondegenerate material.}
\label{f.other_compositions}
\end{figure*}

\begin{figure*}[!htb]
\centering
\includegraphics[width=6.5in]{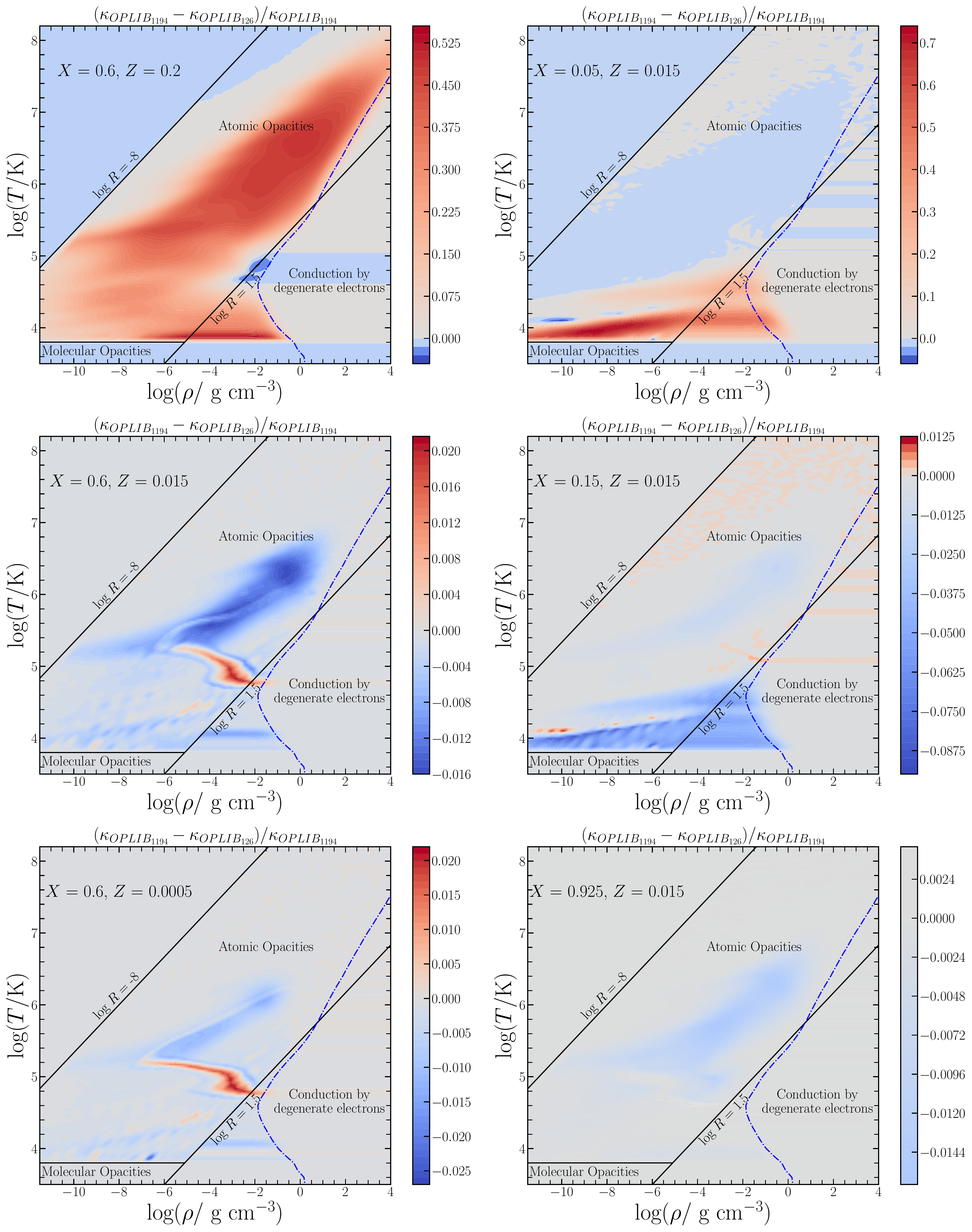} \\
\caption{Relative differences between the 1194 OPLIB and 126 OPLIB opacity table grids using \citet{grevesse_1998_aa} abundances, generated from \MESA's \texttt{kap} module, for six mixtures. The left column shows mixtures with $X$\,=\,0.6 and varying $Z$, and the right column shows mixtures with $Z$\,=\,0.015 and varying $X$.  The OPLIB log($R$)\,=\,$-$8, 1.5 table boundaries are marked with a solid black line.  The approximate location of the Z-dependent transition to an electron conduction dominated opacity is marked with dot-dash blue curve.}
\label{f.panelF}
\end{figure*}

\begin{figure*}[!htb]
\centering
\includegraphics[width=6.5in]{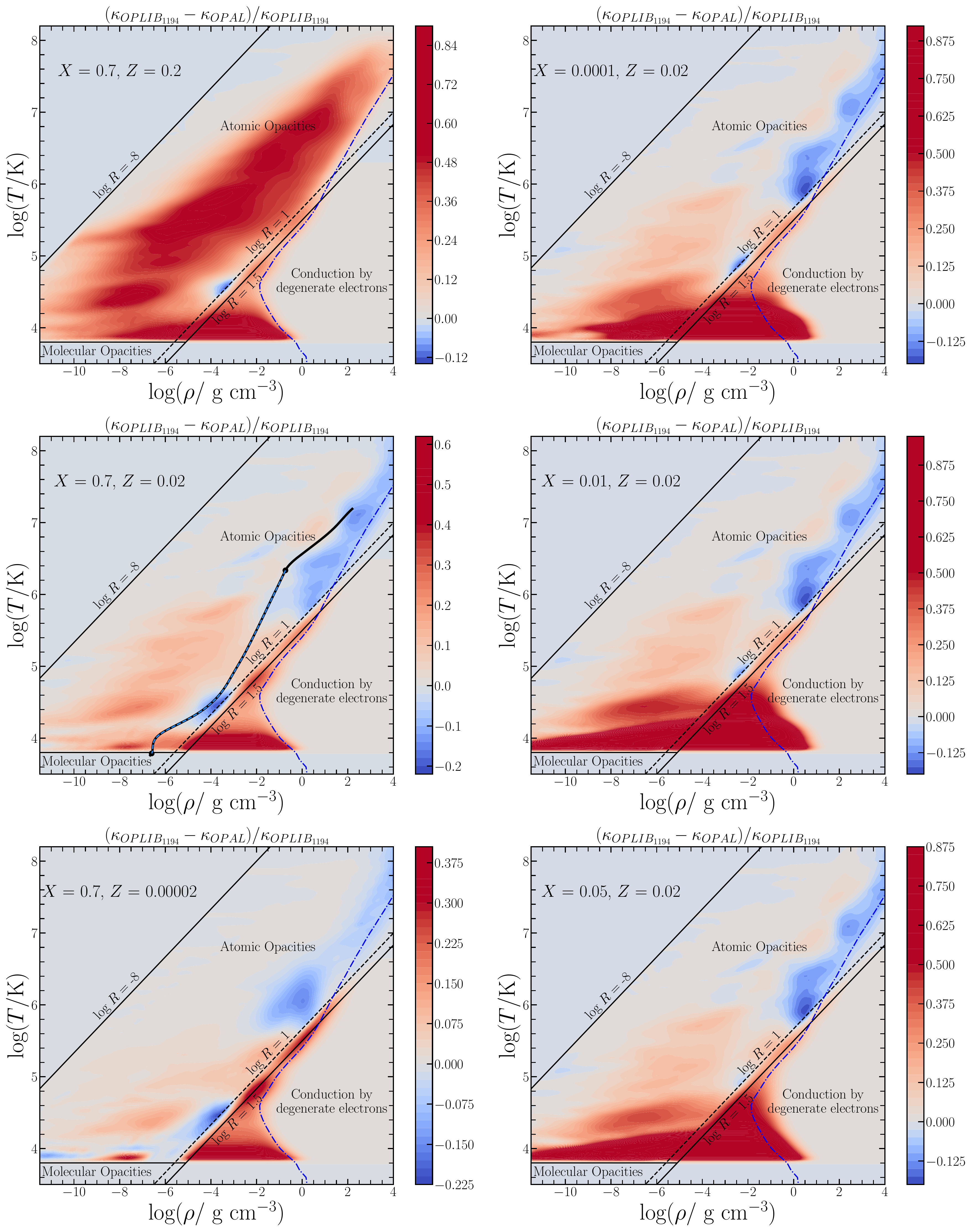} \\
\caption{Relative differences between the 1194 OPLIB and 126 OPAL opacity table grids using \citet{grevesse_1998_aa} abundances, generated from \MESA's \texttt{kap} module, for six mixtures. The left column shows mixtures with $X$\,=\,0.7 and varying $Z$, and the right column shows mixtures with $Z$\,=\,0.02 and varying $X$. The OP/OPAL log($R$)=1 table boundary is marked with a dashed black line, the OPLIB log($R$)\,=\,1.5  table boundary is marked with a solid black line, and  the approximate location of the Z-dependent transition to an electron conduction dominated opacity is marked with dot-dash blue curve.  The thick black curve for the $X$=0.7, $Z$=0.02 mixture marks the location of a standard solar model; the black dots mark the approximate beginning and end of the solar convection zone (overlaid blue dashed curve). Interpolation across the $X$--$Z$ plane uses cubic splines.}
\label{f.panelD}
\end{figure*}

The middle row of Figure~\ref{f.panelA} shows $\partf{\kappa}{T}$ and the bottom row shows $\partf{\kappa}{\rho}$ for this chemical mixture. The left column shows these quantities as a function of $T$ for different $\rho$, and the right column shows contours of these partial derivatives in the $\rho - T$ plane. Peaks and changes of slope in these four plots correlate with ionization stages 
due to the injection of additional electrons into the plasma. Examples include the onset of light metal ionization at low $T$, and the ionization stages of H and He at larger $T$. The sign and magnitude of the partial derivative features are relevant for variable stars driven by the ionization stages of different elements (e.g., He ionization and RR Lyrae variables).

Figure~\ref{f.other_compositions} compares the $\kappa$ generated from OPLIB radiative opacities and \MESA's \texttt{kap} module at log($\rho$/g\,cm$^{-3}$) =\,$-$2, $-$6, and $-$10 for five compositions: the same $X$ and $Z$ as in Figure~\ref{f.panelA}, pure H, a solar $Z$ with the remainder H, pure He, and a solar $Z$ with the remainder He. We first analyze the three H-rich mixtures and then the two He-rich mixtures.

All three H-rich mixtures show a peak $\kappa$ that shifts towards higher $T$ as $\rho$ increases, reflecting the half-ionization curve of H discussed for Figure~\ref{f.panelA}. A difference occurs in $\kappa$ between these three H-rich mixtures at log($T$/K)\,$\lesssim$\,3.8, reaching a three orders of magnitude difference for log($\rho$/g\,cm$^{-3}$)\,=\,$-$2 at log($T$/K)\,=\,3.5. Here, the lack of electrons from metals with relatively low ionization potentials in the pure H composition (blue curve) delays the onset of a dominant H$^{-}$ opacity until H ionizes. Another difference occurs at log($T$/K)\,$\gtrsim$\,5.0 for all three densities, where the lack of electrons from Fe group atoms in the pure H composition (blue curve) suppresses the bound-free and free-free opacities relative to the two mixtures with a solar compliment of Fe.

Helium contributes four main opacity sources: He$^-$ free–free \citep{somerville_1965_aa,john_1994_aa}, He$^{+}_{2}$ bound–free and free-free \citep{ignjatovic_2009_aa}, He Rayleigh scattering \citep{iglesias_2002_aa,rohrmann_2018_aa}, and triple-He collision \citep{kowalski_2014_aa,blouin_2019_aa}. The two He-rich mixtures in Figure~\ref{f.other_compositions} (orange curves) have opacities that are usually about an order of magnitude smaller for a given $T$ and $\rho$ than the opacities of the H-rich mixtures before electron scattering dominates. Below log($T$/K)\,$\lesssim$\,4.2, the presence of free electrons from the ionization of trace metals with relatively low ionization potentials introduces He$^-$ free–free electron scattering and He$^+_2$ bound–free absorption that increases $\kappa$ by several orders of magnitude in the He composition with solar $Z$  \citep{saumon_2022_aa}. Above log($T$/K)\,$\gtrsim$\,5.0 for all three densities, the lack of electrons from Fe group atoms in the pure He composition (dark orange curve) reduces the bound-free and free-free opacities relative to the mixture with a solar Fe abundance (light orange curve). The $\kappa$ curves for log($\rho$/g\,cm$^{-3}$)\,=\,$-$2, $-$6 show undulations due to He$^{0}$ in excited states, H$^{0}$ in its ground state and He$^{+}$ in excited states, and He$^{+}$ in its ground state \citep{seaton_1994_aa}. Like the three H-rich mixtures, both He-rich mixtures also show a peak $\kappa$ that shifts towards higher $T$ as $\rho$ increases. An approximate expression for the half-ionization curve for a pure atomic He mixture is
\begin{equation} \label{eq.half_he}
\rho = 4 \cdot 8.02\times10^{-9} \ T^{3/2} \ \exp(-2.853\times10^5 / T)
\ .
\end{equation}
The half-ionization points for pure atomic H (Equation~\ref{eq.half_h}) and pure atomic He (Equation~\ref{eq.half_he}) mixtures are shown by the black circles in Figure~\ref{f.other_compositions}. These approximations broadly reflect the trend, but miss the $\kappa$ peaks due to missing physics in the approximations, for example the abundance of H$^-$ in a hydrogen mixture.

\subsection{Verification of the OPLIB opacities}\label{s.verification}

Here we illustrate some differences between the 1194 OPLIB opacity tables, a degraded set of 126 OPLIB opacity tables, and the 126 OPAL opacity tables. Figure~\ref{f.panelF} shows the relative opacity differences between the OPLIB 1194 table grid (denoted $\kappa_{OPLIB_{1194}}$) and the OPLIB 126 table grid (denoted $\kappa_{OPLIB_{126}}$) for six mixtures in the $\rho-T$ plane. The six mixtures are chosen so that they lie between or outside the 126 OPLIB table grid in the $X$--$Z$ plane. For constant $X = 0.6$ across $Z = 0.0005,0.015,0.2$, we find interpolation over the 1194 table grid results in $\approx$\,2--2.5\% change in opacity over the 126 table grid for $Z\leq0.1$. Since the 126 OPLIB opacity tables only reach $Z$\,=\,0.1, \MESA\ returns the opacity value at $Z$\,=\,0.1 for $Z$\,$>$\,0.1. For $Z$\,=\,0.2 in the top left panel of Figure~\ref{f.panelF}, the 1194 OPLIB tables provide improvements up to $\approx$\,50\% in \MESA.  The right column of Figure~\ref{f.panelF} shows the opacity differences at $Z = 0.015$ across $X = 0.05,0.15,0.925$. We find interpolation over the 1194 OPLIB tables show a $\approx$\,2\% change in the opacity at larger H mass fractions and up to 75\% change in the opacity at smaller H mass fractions in the temperature range 3.8\,$\lesssim$\,$\log$T\,$\lesssim$\,4.8 due to the ionization of H and He. These differences are a direct result of the enhanced table density between 0\,$\leq X$\,$\leq$\,0.1.  

Figure~\ref{f.panelD} shows the opacity differences between the OPLIB 1194 tables and the OPAL 126 tables for six mixtures in the $\rho-T$ plane. In regions between the atomic radiative opacities and the conductive opacities with no table coverage, there can be large differences since \MESA\ uses the values from the log($R$) table edge and then extends/mixes these radiative opacities with the conductive opacities. Overall, we find differences between the 1194 OPLIB tables and the 126 OPAL tables extend from $\simeq$\,20\% at $X$=0.7, $Z$\,=\,0.02 to upwards of 40--80\% for other mixtures, especially at low $T$ and moderate $\rho$. In the case a solar model profile, shown in the left--middle panel at approximately solar metallicity, $X = 0.7$ and $Z = 0.02$, OPLIB opacity tables consistently predict a higher opacity at the base of the convective envelope and lower opacity in the Solar core at otherwise identical conditions.

\section{Standard Solar Models}\label{s.models}
In this section we detail the construction of standard solar models and their input physics. We quantify the resulting differences in their internal structure and compare with observational helioseismic and neutrino flux constraints. Standard solar models have been previously calculated using OPLIB opacities in \citet{colgan_2016_aa,guzik_2016_aa,raecke_2022_aa} and OPLIB-L opacities by \citet{neuforge-verheecke_2001_ab}. We construct standard solar models with four differing photospheric estimates of the solar heavy element abundance: $Z/X$~$= 0.0181$ \citep[AGSS09,][]{asplund_2009_aa}, $Z/X$~$= 0.0229$ \citep[GS98,][]{grevesse_1998_aa}, $Z/X$~$= 0.0187$ \citep[AAG21,][]{asplund_2021_aa}, and $Z/X$~$= 0.0225$ \citep[MB22,][]{magg_2022_aa}. We compare solar models computed with OPLIB opacities to the default OP and OPAL opacity tables provided by \MESA. We broadly find that standard solar models produced with OPLIB opacities have systematically lower core opacities and temperatures, and higher core densities resulting in lower neutrino fluxes and markedly different helioseismology than OP/OPAL models. We also find that standard solar models computed with higher metallicity mixtures (GS98 or MB22) are in better agreement with helioseismic and neutrino constraints than low metallicity mixtures (AGSS09 or AAG21) regardless of the adopted opacity table.

\vspace{0.2in}

\subsection{Input Physics}\label{s.input_physics}
We use \MESA\ version r22.11.1 to construct our stellar models \citep{paxton_2011_aa,paxton_2013_aa,paxton_2015_aa,paxton_2018_aa,paxton_2019_aa,jermyn_2023_aa}. Each star is modeled as a single, non-rotating, non-mass losing, solar metallicity object. We use the built-in \MESA\ nuclear reaction network \texttt{mesa\_28}. Relatively large nuclear networks are required to fully capture the energy generation rate \citep{farmer_2016_aa}, and thus for example the neutrino luminosity \citep{farag_2023_aa}. The current defaults for nuclear reaction rates are described in Appendix A.2 of \citet{paxton_2019_aa}.  

Rates are taken from a combination of NACRE \citep{angulo_1999_aa} and the Joint Institute for Nuclear Astrophysics REACLIB library (default version, dated 2017-10-20) \citep{Cyburt_2010_ab}, supplemented with NACRE II reaction rates provided in \citet{xu_2013_aa}. We adopt the weak-decay rate for $^{7}$Be from \citet{simonucci_2013_aa}. 
Note there is a long literature on the sensitivity of solar model neutrino fluxes to uncertainties in the nuclear physics \citep{bahcall_1996_aa,deglinnoccenti_1997_aa,bahcall_2004_aa,bahcall_2006_aa,haxton_2008_aa,adelberger_2011_aa,haxton_2013_aa,vissani_2019_aa,villante_2021_aa,bellinger_2022_aa}.
 The \MESA\ screening corrections are from \citet{chugunov_2007_aa}, which includes a physical parameterization for the intermediate screening regime and reduces to the familiar weak \citep{dewitt_1973_aa, graboske_1973_aa} and strong \citep{alastuey_1978_aa,itoh_1979_aa} limits at small and large values of the plasma coupling parameter.  All the weak reaction rates are based (in order of precedence) on the tabulations of \citet{langanke_2000_aa}, \citet{oda_1994_aa}, and \citet{fuller_1985_aa}. 
  
 We use \MESA's default equation of state \citep{jermyn_2023_aa}, where a standard solar model's $\rho - T$ profile lies in the domain of FreeEOS calculated with a \citet{grevesse_1998_aa} composition. The atmosphere is modeled using the solar-Hopf relation analytic fit by \citet{ball_2021_aa} to the Hopf function for the solar simulation by \citet{trampedach_2014_aa}.

We adopt the \MESA\ default  time-dependent convection model which reduces to the Cox MLT model \citep{cox_1968_aa} on long timescales typical of a solar model \citep{jermyn_2023_aa}. We adopt the Ledoux criterion for convective stability, and we adopt a semiconvection efficiency coefficient of $\alpha$\,=\,0.1. We use the convective premixing routine \citep{paxton_2019_aa} to determine the location of convective boundaries, and we do not include the effects of convective overshooting, rotational deformation, or the effects of rotational mixing. We include the effects of element diffusion for every isotope included in our nuclear network by solving the unmodified Burgers equations in cgs units and including the heat flow vector terms \citep{paxton_2018_aa}. Each model is self-consistently evolved with the hydrodynamics such that a radial velocity variable is present throughout the evolution.

\begin{deluxetable*}{lcccccccc}[!htb]
  \tablenum{1}
  \tablecolumns{8}
  \tablewidth{\columnwidth}
  \tablecaption{Solar Calibration Parameters and properties \label{tab:table1}}
  \tablehead{
      \colhead{Component} & \colhead{$X$\textsubscript{0} } & \colhead{$Y$\textsubscript{0}} & \colhead{$Z$\textsubscript{0} } & \colhead{$\alpha_{mlt}$ } & \colhead{($Z/X$)\textsubscript{surf}} & \colhead{$L_{\nu,\odot}/L_{\gamma,\odot}$} & \colhead{$R$\textsubscript{cz,b}/$R_{\odot}$} & \colhead{$Y$\textsubscript{surf} }}
  \startdata
  \textbf{Currently in \code{MESA}} \\
	OPAL\_126 GS98 & 0.70964 & 0.27181 & 0.01855 & 1.799 & 0.0229 & 0.024181 & 0.718 & 0.24423\\
 	OPAL\_126 AGSS09 & 0.71990  & 0.265125 & 0.01498 & 1.778 & 0.0181 & 0.023909 & 0.726 & 0.23697  \\ 
 	OP\_126 GS98 & 0.71114 & 0.27028 & 0.01857 & 1.810 & 0.0229 & 0.024133 & 0.718 & 0.24298 \\
 	OP\_126 AGSS09 & 0.72009 & 0.26490 & 0.01500 & 1.783 & 0.0181 & 0.023882 & 0.726 & 0.23674 \\
  \hline{}
  \textbf{From this Work} \\
   	OPLIB-L\_126\_50T GS98  & 0.71156 & 0.26975 & 0.01869 & 1.9054 & 0.0229 & 0.024150 & 0.7235 & 0.24186 \\
    OPLIB\_126\_50T GS98  & 0.71379 & 0.26762 & 0.01859 & 1.9494 & 0.0229 & 0.023977 & 0.7167 & 0.24108 \\
 	OPLIB\_126 GS98  & 0.71202 & 0.26944   & 0.01854 & 1.9442 & 0.0229 & 0.024010 & 0.7166 & 0.24275 \\
 	OPLIB\_126 AGSS09 & 0.72564 & 0.25931  & 0.01504 & 1.9403 & 0.0181 & 0.023679 & 0.7230 & 0.23258 \\ 
    OPLIB\_126 AAG21 & 0.72654 & 0.25795   & 0.01551 & 1.9140 & 0.0181 & 0.023658 & 0.7214 & 0.23172  \\
 	OPLIB\_126 MB22  & 0.71480 & 0.26694   & 0.01826 & 1.9339 & 0.0225 & 0.023986 & 0.7151 & 0.24076 \\
    OPLIB\_1194 GS98 & 0.71191 & 0.26955   & 0.01854 & 1.9437 & 0.0229 & 0.024012 & 0.7166 & 0.24284 \\
 	OPLIB\_1194 AGSS09 & 0.72607 & 0.25886 & 0.01507 & 1.9350 & 0.0181 & 0.023670 & 0.7248 & 0.23193 \\
 	OPLIB\_1194 AAG21  & 0.72684 & 0.25762 & 0.01554 & 1.9095 & 0.0181 & 0.023651 & 0.7225 & 0.23122 \\
    OPLIB\_1194 MB22 & 0.71469 & 0.26705   & 0.01826 & 1.9331 & 0.0225 & 0.023988 & 0.7159 & 0.24082 \\
   \hline{}
\textbf{\citet{vinyoles_2017_aa}} \\   
OP\_126 GS98  &  0.7095 & 0.2718 & 0.0187 & 2.18 & 0.0229 & - & 0.7116 & 0.2426 \\
OP\_126 AGSS09met & 0.7238 & 0.2613 & 0.0149 & 2.11 & 0.0187 & - & 0.7223 & 0.2317 \\
\textbf{\citet{magg_2022_aa}} \\   
OP\_126 GS98      & 0.7095 & 0.2718 & 0.0187 & - & 0.0229 & - & 0.7122 & 0.2425  \\
OP\_126 AGSS09met & 0.7237 & 0.2614 & 0.0149 & - & 0.0178 & - & 0.7231 & 0.2316  \\
OP\_126 AAG21     & 0.7207 & 0.2638 & 0.0155 & - & 0.0187 & - & 0.7197 & 0.2343  \\
OP\_126 MB22      & 0.7090 & 0.2734 & 0.0176 & - & 0.0225 & - & 0.7123 & 0.2439  \\
\hline{}
    Observation$^{a}$ & -   & -     & -      &  -  & - &   - & 0.713\,$\pm$\,0.001 & 0.2485\,$\pm$\,0.0035   
  \enddata
  \tablenotetext{a}{The helioseismic derived 
radius at the bottom of the convective zone, $R$\textsubscript{cz,b}, 
and surface He mass fraction, $Y$\textsubscript{surf},
are from \citet{basu_1997_aa} and \citet{basu_2004_aa}.
}
\end{deluxetable*}

\begin{figure}[!htb]
\centering
\includegraphics[width=3.38in]{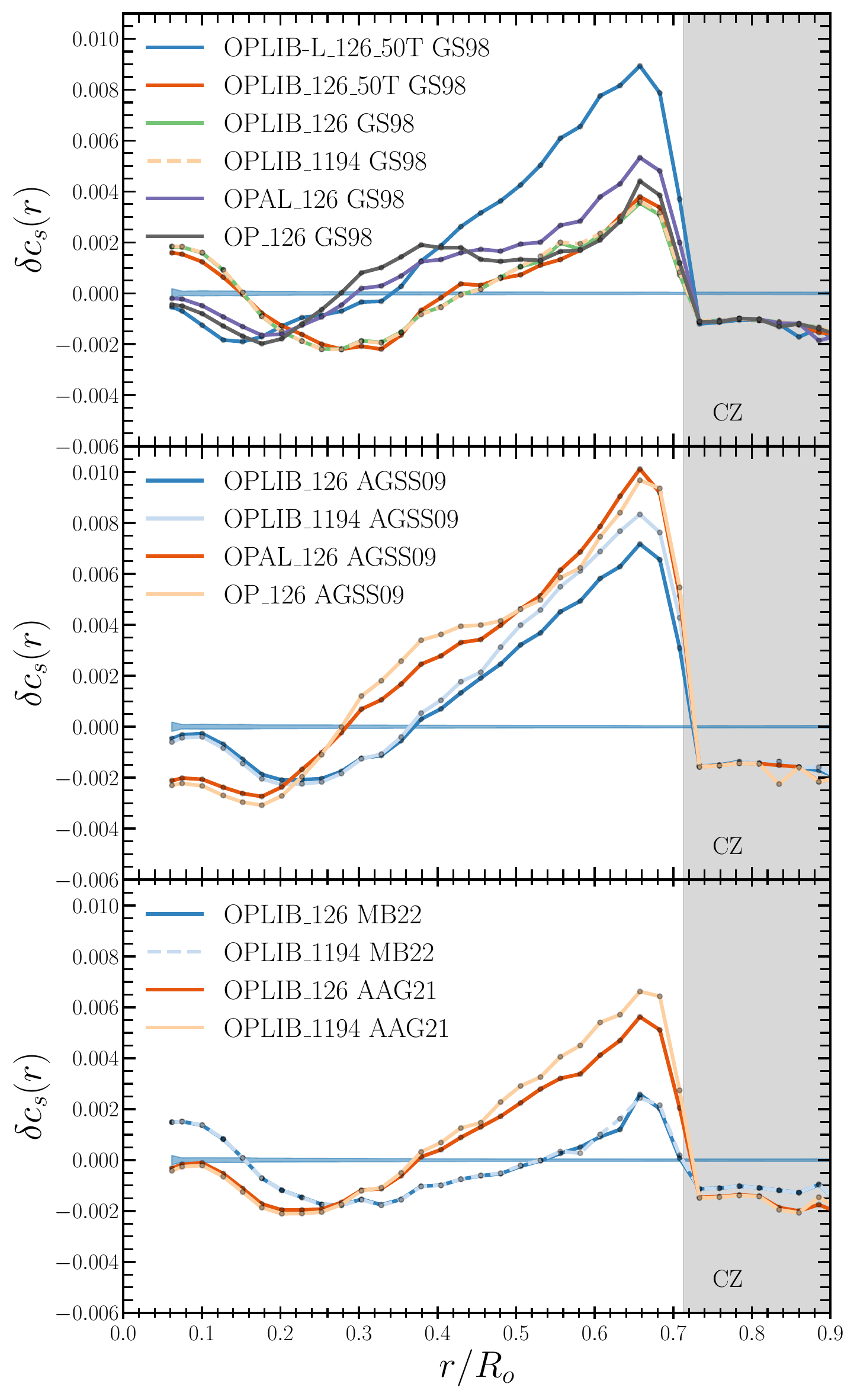} \\
\caption{
Fractional sound speed differences $\delta c_{s}$ $=$ $(c_{\rm obs}$ - $c_{s}(r)) /c_{s}(r)$ 
between values predicted by a calibrated \MESA\ standard solar model $c(r)$
and $c_{\rm obs}$ values inferred from helioseismic data  \citep{basu_2009_aa}.
The 1$\sigma$ observational uncertainties are shown as the blue bands at ordinates of zero.
Top, middle, and bottom panels are for GS98, AGSS09, and MB22+AAG21 mixtures respectively.
Black circles mark locations where $\delta c_{s}$ is evaluated. 
Gray bands show the convective regions, labeled CZ.
}
\label{f.c_diff1}
\end{figure}

\subsection{Standard Solar Model Calibrations}\label{s.solar_norm}

We perform solar model calibrations to generate standard solar models and compare our results with present day helioseismic and neutrino observation data.
We iterate on differences between the final model at $t_{\sun}$\,=\,4.568~Gyr \citep{bouvier_2010_aa}
and the solar radius, $R_{\sun}= 6.957 \times 10^{10}$ cm, 
solar luminosity, $L_{\gamma,\sun}= 3.828 \times 10^{33}$ erg s$^{-1}$ \citep{prsa_2016_aa}, 
and surface heavy element abundance $Z/X$.  

We use the built-in \MESA\ simplex module to iteratively vary 
the mixing-length parameter, $\alpha$, and the initial composition $X$, $Y$, and $Z$. This calibration is performed for four estimates of the heavy element abundance at
the surface of the Sun: $Z$/X~$= 0.0181$ \citep[AGSS09,][]{asplund_2009_aa}, 
$Z/X$~$= 0.0229$ \citep[GS98,][]{grevesse_1998_aa}, $Z/X$~$= 0.0187$ \citep[AAG21,][]{asplund_2021_aa}, and $Z/X$~$= 0.0225$ \citep[MB22,][]{magg_2022_aa}. 

Calibrated parameters are listed in Table \ref{tab:table1}. The solar models are calculated using OPAL \citep{iglesias_1996_aa},  Opacity Project (OP) \citep{badnell_2005_aa}, OPLIB, and OPLIB-L (see Section~\ref{s.atomic_code}) Rosseland-mean radiative opacities.

Each of the 16 stellar models calculated in this work has approximately 5000 mass zones. Each model takes approximately 1200 timesteps. Each model takes approximately 2-4 hours to run on a 12--16 core machine. Each solar model is run roughly 150-200 times to calibrate, taking roughly 2--4 weeks to complete. The \MESA\ files to reproduce our models, along with the python scipts to reproduce all of our plots are available at 
\dataset[http://doi.org/10.5281/zenodo.10798600]{http://doi.org/10.5281/zenodo.10798600}.

\subsection{Helioseismic Observables}\label{s.helio}
Helioseismic inversions of solar models in combination with solar oscillation data provide observational estimates of quantities such as the radius of the solar convection zone base 
\citep{christensen-dalsgaard_1991_aa,kosovichev_1993_aa,basu_1997_aa}, the convection zone helium mass fraction \citep{vorontsov_1991_aa,dziembowski_1991_aa,antia_1994_aa,basu_1995_aa,richard_1998_aa}, radial profiles of sound speed and density \citep{christensen-dalsgaard_1985_aa,antia_1994_ab} two-dimensional profiles of the rotational velocity \citep{thompson_1996_aa,brown_1987_aa}, and other quantities such the Ledoux discriminant profile \citep{buldgen_2020_aa}. In this section we compare inferred helioseismic quantities from \citet{basu_1997_aa,basu_2004_aa,basu_2009_aa} with those predicted by standard solar models with (GS98, AGSS09, AAG21, and MB22) abundances and (OPLIB, OPAL, and OP) opacity tables.

\begin{figure*}[!htb]
\centering
\includegraphics[width=7in]{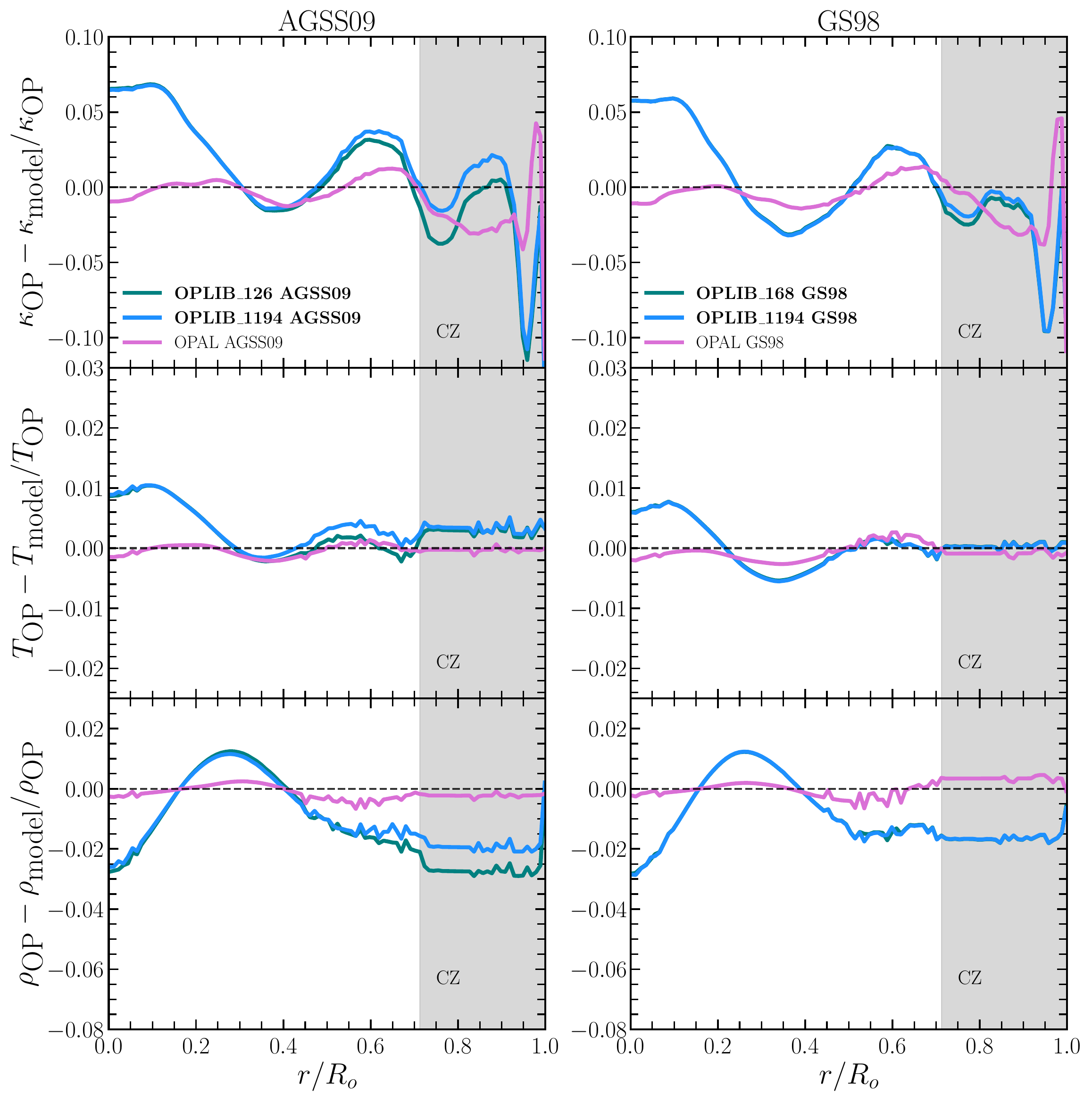} \\
\caption{ Opacity (top), temperature (middle), and density (bottom) differences between standard solar models normalized to a standard solar model computed with OP opacities \citep{badnell_2005_aa}. The left column is for the \citet{asplund_2009_aa} mixture and the right column is for the \citet{grevesse_1998_aa} mixture. New results are labeled in bold font. We use a smoothing floor to reduce the noise generated by taking finite differences between small changes in radii.
}
\label{f.norm_diff}
\end{figure*}

\begin{figure}[!htb]
\centering
\includegraphics[width=3.38in]{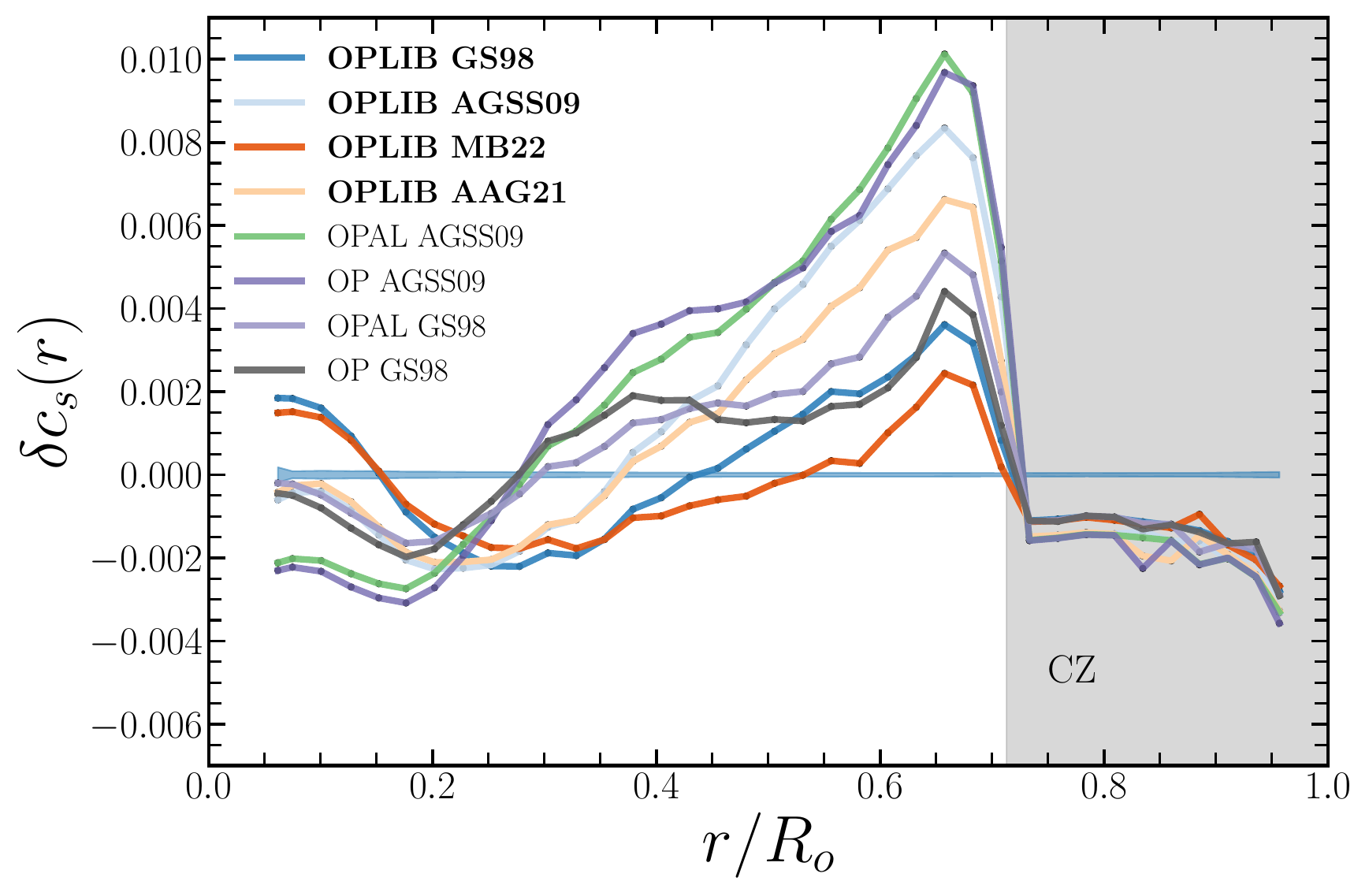} \\
\includegraphics[width=3.38in]{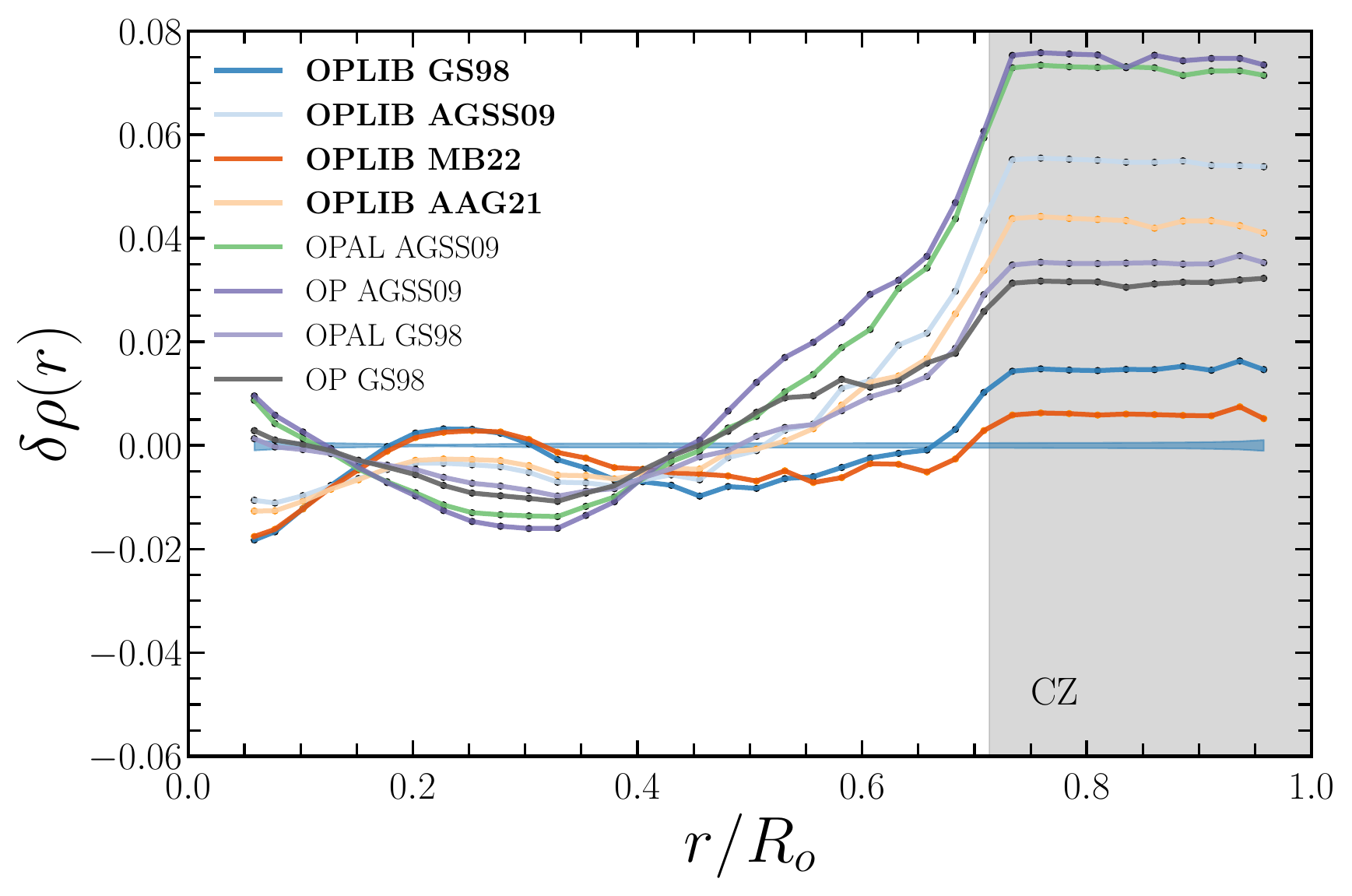} \\
\caption{
Fractional sound speed and density differences, 
$\delta c_{s}$ $=$ $(c_{\rm obs}$ - $c_{s}(r)) /c_{s}(r)$ and $\delta \rho$ $=$ $(\rho_{\rm obs}$ - $\rho(r)) /\rho(r)$, 
between the values predicted by a calibrated \MESA\ standard solar model, $c_{s}(r)$ and $\rho(r)$, 
and the  $c_{\rm obs}$ and $\rho_{\rm obs}$ values inferred from helioseismic data \citep{basu_2009_aa}.
The 1$\sigma$ observational uncertainties are shown as the blue bands at ordinates of zero. 
Black circles mark locations where $\delta c_{s}$ and $\delta \rho$ are evaluated. 
Gray bands shows the convective regions, labeled CZ. New results are labeled in bold font.
}
\label{f.c_diff2}
\end{figure}

Calculated helioseismic quantities are shown in Table~\ref{tab:table1} for each set of opacity tables considered in this work and from \citet{vinyoles_2017_aa} and \citet{magg_2022_aa}. Since we have evolved these solar models with otherwise identical input physics, disagreements between models of identical chemical mixtures should arise from differences in the treatment of atomic opacities. In Appendix~\ref{appendix.C} we highlight the necessity of adopting cubic interpolation across $X$--$Z$ in \MESA\ to reproduce similar helioseismic results found in the other works.

The first column of Table~\ref{tab:table1} uses the notation ``Source\_N", where ``Source" is OPLIB, OPAL, or OP and ``N" is either 126 or 1194 signifying the number of individual tables. With two exceptions, all the models shown in Table~\ref{tab:table1} use an individual $\kappa$ table resolution described in Section~\ref{s.tables}. The two exceptions are entry OPLIB\_126\_50T which refers to a calibrated solar model computed with 126 individual OPLIB tables at a table resolution of 50 temperature points, and entry OPLIB-L\_126\_50T which uses the older OPLIB-L tables at a table resolution also of 50 temperature points.

Table~\ref{tab:table1} also lists the initial abundances $X$\textsubscript{0}, $Y$\textsubscript{0} and $Z$\textsubscript{0} along with the calibrated dimensionless mixing length parameter $\alpha_{mlt}$, the surface metal to H abundance ratio ($Z/X$)\textsubscript{surf}, the neutrino luminosity $L_{\nu,\odot}/L_{\gamma,\odot}$, the radius of the convection zone base $R$\textsubscript{cz,b}/$R_{\odot}$, and the surface helium mass fraction $Y$\textsubscript{surf}. We find both OP GS98 and AGSS09 \MESA\ solar models agree well with the results found in both \citet{vinyoles_2017_aa} and \citet{magg_2022_aa}, both of which use the GARSTEC \citep[GARching STEllar Code,][]{weiss_2008_aa} with the most up to date physical prescriptions from the ``B16" solar models. These \MESA models adopts the photospheric AGSS09 abundance mixtures. \citet{vinyoles_2017_aa} and \citet{magg_2022_aa} adopt the AGSS09 photospheric abundance mixture supplemented by meteoric abundances for refractory elements (labeled ``AGSS09met" in Table~\ref{tab:table1}). Some of the difference between the \MESA\ OP AGSS09 and the \citet{vinyoles_2017_aa} OP AGSS09met models can be attributed to these composition differences. Overall, OPLIB standard solar models produce a smaller $R$\textsubscript{cz,b}/$R_{\odot}$ and smaller $Y$\textsubscript{surf} as compared to OP or OPAL solar models for similar compositions.

Figure~\ref{f.c_diff1} shows the fractional difference in sound speed, $\delta c_{s}$, for each calibrated solar model in Table~\ref{tab:table1} with respect to the helioseismic sound speeds inferred in \citet{basu_2009_aa}.  The OPLIB-L\_126\_50T model in the top panel (blue curve) agrees more with OP and OPAL models (gray and purple curves respectively) in the stellar core for $r/R_{o}$~$\lesssim$~0.4, but has a significantly larger $\delta c_{s}$ as compared to the OPLIB, OP, or OPAL solar models for $r/R_{o}$~$\gtrsim$~0.4. The OPLIB\_126\_50T model (red curve) shows similar sound speeds to OPLIB\_126 (green curve) showing that the improved individual table resolution in the OPLIB\_126 model does not have a measurable large impact on $\delta c_{s}$.  The three OPLIB models have higher sound speeds compared to the OP, OPAL, and OPLIB-L models for $r/R_{o}$~$\lesssim$~0.2, and lower sound speeds for 0.2~$\lesssim$~$r/R_{o}$~$\lesssim$~0.4. In the envelope for $r/R_{o}$~$\gtrsim$~0.4 the three OPLIB models are closer to the OP and OPAL models, yet show a difference in $\delta c_{s}$. 

The AGSS09 mixture in the middle panel show improvement in $\delta c_{s}$ with OPLIB opacities (dark and light blue curves). The AAG21 and MB22 mixtures with OPLIB opacities in the bottom panel show overall better agreement with the inferred solar sound speed than the GS98 and AGSS09 mixtures.

Near the peaks in $\delta c_{s}$, below the convection zone base, the 1194 OPLIB models have a larger $\delta c_{s}$ than the 126 OPLIB models for the AGSS09 and AAG21 mixtures (second and third panels) due to a lower opacity from the more accurate interpolations offered by the 1194 table set. The difference is more pronounced for the AGSS09 and AAG21 mixtures due to their lower metallicity, with $Z_{0}$~$\sim$~0.015 as opposed to $Z_{0}$~$\sim$~0.018-0.019 in the GS98 and MB22 mixtures where using 1194 or 126 tables has little consequence. A metallicity of $Z_{0}$~$\sim$~0.015 lies between the $Z$~$\sim$~0.01 and $Z$~$\sim$~0.02 tabulations of the default 126 table set. The new 1194 table set provides tables at $Z$~$\sim$~0.014,~0.015,~0.016.

Figure~\ref{f.norm_diff} compares the $\kappa$, $T$, and $\rho$ profiles of standard solar models evolved with GS98 and AGSS09 mixtures. The differences are normalized to standard solar models computed with OP opacities \citep{badnell_2005_aa}, as they are commonly adopted standard in solar modeling studies. Models evolved with OPLIB opacities (blue curves) have $\sim$\,6--7\% lower core opacity than OP (dashed horizontal line) over $r/R_{o}$\,$\lesssim$\,0.15, while the model with OPAL opacity (purple curve) is within $\sim$\,1\% of OP across the entire radiative core. Between 0.15~$\lesssim$~$r/R_{o}$~$\lesssim$~0.4 the OPLIB normalized opacity difference decreases to a local minimum near $r/R_{o}$~$\sim$~0.4, and increases to $\sim$\,$3$\%  at $r/R_{o}$\,$\sim$\,0.65 below the convection zone base. 

Differences in $\kappa$ arising from the number of OPLIB tables adopted are visible in Figure~\ref{f.norm_diff} at $r/R_{o}$~$\gtrsim$~0.7 for the GS98 and $r/R_{o}$~$\gtrsim$~0.5 for AGSS09 mixtures. Overall, we find differences in the opacity table source (OP, OPAL, OPLIB) result in solar model opacity differences of $\approx 8 \%$ and $\approx 15 \%$ at the bottom and top of the solar convection zone, and up to $\approx 7 \%$ in the solar core. It should be noted that only a small portion of the total luminous flux in a convective envelope is carried by radiation, hence its structure is primarily determined by the EOS not the opacity. However opacity differences are important near the bottom and top of the solar convection zone, because the convective boundary location is where adiabatic and radiative temperature gradients are comparable to one another. The opacity is important in the shallow super-adiabatic layers, as it sets the entropy gradient for the convection zone as a whole, as well as the radiative atmosphere. In more massive, hotter stars than the Sun, these opacity differences can become very important because their envelopes can become radiation dominated and are not always fully convective. 

Standard solar models with the OPAL and OP tables have $T$ and $\rho$ profiles that are  within 0.1\% of one another, while models with the OPLIB opacities that have 0.8--1\% lower $T$ and 2.5--3\% higher $\rho$ in the core. Outside of 0.2~$\leq$~$r/R_{o}$~$\leq$~0.4 OPLIB solar models show $\gtrsim$\,1--2\% lower $\rho$ than OP/OPAL solar models. 

Figure~\ref{f.c_diff2} shows the fractional difference in sound speed and density between calibrated solar models and inferred helioseismic values \citep{basu_2009_aa}.
Models using the 1194 OPLIB table set are compared to models using the OP and OPAL tables across different initial compositions. Overall, the MB22 model shows the smallest $\delta c_{s}$ and $\delta \rho$ differences. This suggests a high-$Z$ solution to the solar model problem \citep{guzik_2008_aa,guzik_mussack_2010_aa,serenelli_2009_aa,salmon_2021_aa}. Recent helioseismic determination of the solar mass-fractions by \citet{buldgen_2023_aa,buldgen_2023_ab} favors low-$Z$ mixtures such as AAG21. However, these two results are not necessarily in tension as the model sets are constructed and calibrated differently.

\begin{figure*}[!htb]
\centering
\includegraphics[width=6.5in]{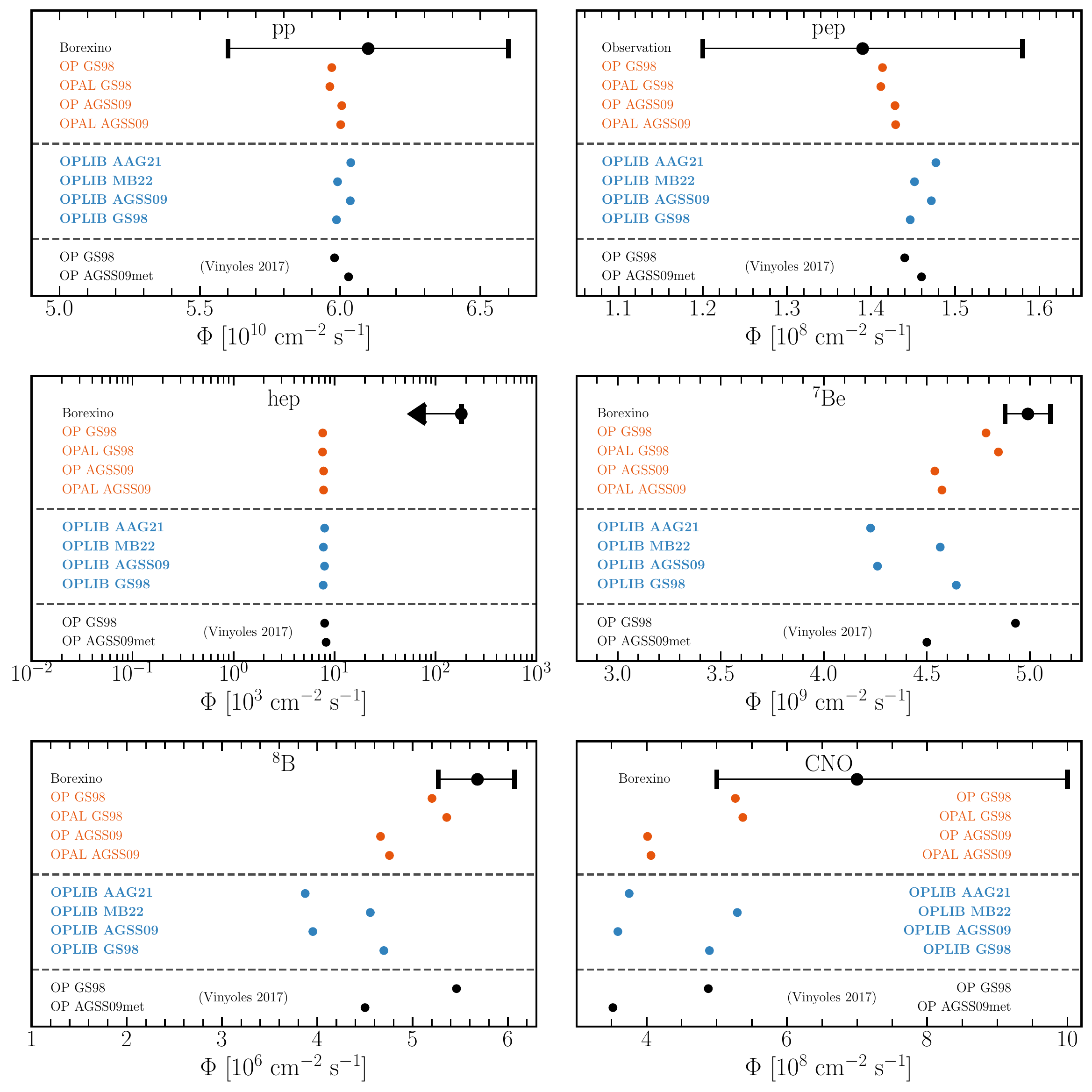} \\
\caption{Neutrino fluxes and uncertainties compared to solar neutrino observations from the Borexino Collaboration \citep{bellini_2011_aa} as presented in \citet{haxton_2013_aa}, \citet{villante_2014_aa}, and  \citet{kumaran_2021_aa}. Standard solar model results using previous opacity sets are shown in orange, results using the new 1194 OPLIB opacity tables are shown in blue, and results from \citet{vinyoles_2017_aa} are shown in black. See Table~\ref{tab:table2} for the data behind this figure.}
\label{f.neutrinos}
\end{figure*}

\vspace{0.3in}

\subsection{Solar Neutrino Fluxes}\label{s.neutrino}
Neutrinos are produced during H-burning on the main-sequence
from the proton-proton chain reactions
p(p,e$^+$$\nu_e$)$^2$H, 
p(e$^-$p,$\nu_e$)$^2$H, 
$^3$He(p,e$^+$$\nu_e$)$^4$He, 
$^7$Be(e$^-$,$\nu_e$)$^7$Li, 
$^8$B(,e$^+$$\nu_e$)$^8$Be,
and the CNO cycle reactions
$^{13}$N(,e$^+$$\nu_e$)$^{13}$C, 
$^{13}$N(e$^{-}$,$\nu_e$)$^{13}$C,
$^{15}$O(,e$^+$$\nu_e$)$^{15}$N, 
$^{15}$O(e$^{-}$,$\nu_e$)$^{15}$N,
$^{17}$F(,e$^+$$\nu_e$)$^{17}$O, 
$^{17}$F(e$^{-}$,$\nu_e$)$^{17}$O,
$^{18}$F(,e$^+$$\nu_e$)$^{18}$O,
where electron capture reactions on CNO nuclei are included \citep{stonehill_2004_aa}. 
The neutrino flux in the solar interior is strongly dependent on the
core $T$ \citep[see][]{bahcall_1996_aa}. Standard solar models
that accurately predict the $T$ profile of the solar core should also 
generate comparable neutrino fluxes to solar neutrino data \citep{farag_2020_aa}.

Neutrino fluxes $\Phi$ from each \MESA\ calibrated solar model and \citet{vinyoles_2017_aa} models are compared to the observations in Figure~\ref{f.neutrinos} and Table~\ref{tab:table2}. $\Phi$(pp), and $\Phi$(pep) are well within 1\,$\sigma$ of their inferred observational value. The fractional variation in $\delta\Phi$(pp),~$\delta\Phi$(pep)~$\propto$~$-$0.9,~$-$1.4~$\delta T$ \citep{villante_2021_aa}is a nearly linear relationship with the  fractional variation in the temperature $\delta T$. Therefore the variation of nearly $\sim$\,$9$\% in the observational value for  $\Phi$(pp) and $\Phi$(pep) allow for a similar level of deviation in these nuclear reaction rates. However, these rates are constrained at the nearly $\sim$\,1\% level \citep{adelberger_2011_aa}, suggesting the current neutrino flux statistics for $\Phi$(pp) and $\Phi$(pep) do not presently constrain the $T$ of the solar core. Our results show similar $\Phi$(pp) regardless off the adopted opacity table and slightly higher neutrino fluxes from $\Phi$(pep) in the OPLIB based models then the OP and OPAL based models. All solar models, regardless of the adopted opacity table, reliably predict the value of $\Phi$(hep) to be roughly an order of magnitude lower than the currently measured upper limit.

The $\Phi$($^{7}$Be) and $\Phi$($^{8}$B) fluxes place the tightest constraints on the solar core $T$  with $\delta\Phi$($^{7}$Be),~$\delta\Phi$($^{8}$B)~$\propto$~$-$11,~$-$24~$\delta T$ \citep{villante_2021_aa}. The lower core $T$ produced by the OPLIB based solar models result in significantly lower $\Phi$($^{7}$Be) and $\Phi$($^{8}$B) versus the OP or OPAL models regardless of initial $Z$. Higher $Z$ solar models systematically produce larger $^{7}$Be and $^{8}$B neutrino fluxes. The metallicity is most important in determining the CNO flux $\Phi$(CNO) which appears to agree only with the latter metallicity GS98/MB22 models regardless of the chosen opacity table. OPLIB based solar models show a slight decrease in $\Phi$(CNO) resulting from their lower core $T$.

Despite many differences in model physics between our \MESA\ models and the B16 standard solar models from \citet{vinyoles_2017_aa}, our neutrino flux results are in good agreement with theirs. Overall, OPLIB opacities produce sufficiently different neutrino flux and helioseismic quantities to warrant the re-investigation of solar models with additional input physics such as rotation, accretion, and other mixing mechanisms which could alter the chemical stratification of these solar models  \citep{guzik_mussack_2010_aa,srwood_2018_aa,zhang_2019_aa,kunitomo_2021_aa,kunitomo_2022_aa}. It is conceivable that the true solar abundances in the core are higher than in the envelope.

\begin{deluxetable*}{lccccccccc}[!htb]
  \tablenum{2}
  \tablecolumns{7}
  \tablewidth{\columnwidth}
  \tablecaption{Solar Neutrino Fluxes \label{tab:table2}}
  \tablehead{
   \colhead{Component} & \colhead{$\Phi$\textsubscript{pp}} & \colhead{$\Phi$\textsubscript{pep}} & \colhead{$\Phi$\textsubscript{hep}} & \colhead{$\Phi$\textsubscript{Be7}} & \colhead{$\Phi$\textsubscript{B8}}  & \colhead{$\Phi$\textsubscript{CNO}}     
   }
  \startdata
	Observed & $6.1\pm (0.5^{+0.3}_{-0.5})$ & $1.39\pm (0.19^{+0.08}_{-0.13})$ & $\leq180$ (90 \% CL)&  $4.99\pm (0.11^{+0.06}_{-0.08})$ & $5.68\pm (^{+0.39}_{-0.41}$$^{+0.03}_{-0.03})$ & $7.0 ^{+3.0}_{-2.0}$ \\
 \hline{}
 \textbf{Currently in \code{MESA}} \\
	OPAL\_126 GS98 & 5.964 & 1.412 & 7.615 & 4.856 & 5.374 & 5.395 \\
 	OPAL\_126 AGSS09 & 6.003& 1.430 & 7.794 & 4.573 & 4.754 & 4.060  \\
 	OP\_126 GS98 & 5.970& 1.414 & 7.641 & 4.796 & 5.221 & 5.291 \\
 	OP\_126 AGSS09 & 6.007 & 1.429 & 7.803 & 4.540 & 4.662& 4.012 \\
   \hline{}
 \textbf{This Work} \\
  	OPLIB\_168 GS98     & 5.987 & 1.447 & 7.719 & 4.640 & 4.692 & 4.892 \\ 
 	OPLIB\_126 AGSS09   & 6.035 & 1.472 & 7.955 & 4.274 & 3.974 & 3.597 \\
   	OPLIB\_126 AAG21    & 6.038 & 1.477 & 7.976 & 4.227 & 3.872 & 3.751 \\
 	OPLIB\_126 MB22     & 5.991 & 1.452 & 7.762 & 4.562 & 4.551 & 5.288 \\
   	OPLIB\_1194 GS98    & 5.987 & 1.447 & 7.717 & 4.643 & 4.697 & 4.896 \\
    OPLIB\_1194 AGSS09  & 6.036 & 1.472 & 7.961 & 4.260 & 3.952 & 3.590 \\
 	OPLIB\_1194 AAG21   & 6.039 & 1.476 & 7.980 & 4.217 & 3.857 & 3.748 \\
   	OPLIB\_1194 MB22    & 5.991 & 1.452 & 7.760 & 4.565 & 4.555 & 5.293 \\
   \hline{}
   \textbf{\citet{vinyoles_2017_aa}} \\
    OP\_126 GS98      & 5.98 & 1.44 & 7.98 & 4.93 & 5.46 &  4.88 \\
    OP\_126 AGSS09met & 6.03 & 1.46 & 8.25 & 4.50 & 4.50 &  3.52 \\
    \enddata
    \tablenotetext{a}{Observations from the Borexino Collaboration \citep{bellini_2011_aa} as presented in \citet{haxton_2013_aa,villante_2014_aa,kumaran_2021_aa}.
Flux scales for $\Phi$ (in cm\textsuperscript{-2} s\textsuperscript{-1}) are: 
$10^{10}$ (pp); 
$10^{8}$ (pep); 
$10^{3}$ (hep); 
$10^{9}$  (Be);  
$10^{6}$  (B);  
$10^{8}$  (CNO).  
}
\end{deluxetable*}
\section{Conclusion}\label{s.conclusion}

We presented highlights of a new set of 1194 Type-1 Rosseland-mean opacity tables for four different metallicity mixtures. These new Los Alamos OPLIB radiative opacity tables are an order of magnitude larger in number than any previous opacity table release, and span regimes where previous opacity tables have not existed. For example, the new set of opacity tables expands the metallicity range to $Z$\,=\,10$^{-6}$ to $Z$\,=\,0.2 which allows improved accuracy of opacities at low and high metallicity. In addition, the table density in the metallicity range $Z$\,=\,10$^{-4}$ to $Z$\,=\,0.1 are enhanced to improve the accuracy of opacities drawn from interpolations across neighboring metallicities.  Finally, there are new opacity tables for hydrogen mass fractions between $X$\,=\,0 and $X$\,=\,0.1 including $X$\,=\,$10^{-2}, 10^{-3}, 10^{-4}, 10^{-5}, 10^{-6}$ that can improve stellar models of hydrogen deficient stars. This larger set of opacity tables result in a $\approx 2-2.5 \%$ improvement across the $X$--$Z$ plane. At $Z> 0.1$ the improvement can reach up to $50 \%$ in MESA. The largest improvement, up to $\approx 70$, is observed between $3.8\leq$~log$T$~$\leq 4.8$ and $0<X<0.1$: the hydrogen poor regime which can be encountered in stellar models with effective gravitational settling. Differences between using the new 1194 OPLIB opacity tables and the 126 OPAL opacity tables range from $\approx 20-80 \%$ across individual chemical mixtures. 

We implemented and verified these new OPLIB radiative opacity tables in \MESA. We found that calibrated solar models produced with these new OPLIB tables agree broadly with previously published helioseismic and neutrino results found in the literature. In hotter and denser regimes, our calculated OPLIB opacity tables yield higher opacities than OP/OPAL. In colder models such as solar models, OPLIB opacity tables predict lower core opacities, lower core $T$, and higher core $\rho$ than OP/OPAL. Overall, we find differences in the opacity table source (OP, OPAL, OPLIB) result in solar model opacity differences of $\approx 8 \%$ and $\approx 15 \%$ at the bottom and top of the solar convection zone, and up to $\approx 7 \%$ in the solar core.  We find that standard solar models produced with the OPLIB opacities produce lower neutrino fluxes and markedly different helioseismology than OP/OPAL models. We also find that standard solar models computed with higher metallicity mixtures (GS98 or MB22) are in better agreement with helioseismic and neutrino constraints than low metallicity mixtures (AGSS09 or AAG21) regardless of the adopted opacity table. We find that the new OPLIB opacity tables do not solve the solar modeling problem when used in standard solar models. This suggests that physical mechanisms other than the atomic radiative opacity should be further investigated in order to solve the solar modeling problem \citep{guzik_2008_aa,guzik_mussack_2010_aa,serenelli_2009_aa,salmon_2021_aa,eggenberger_2022_aa,buldgen_2023_ab}.

We also tested the opacity interpolation schemes adopted in \MESA\ for interpolation across the $X$--$Z$ plane. We find that linear interpolation systematically under predicts the opacity by $3-60\%$ across metallicities and up to $10 \%$ in solar models, as compared to cubic interpolation. We find \MESA\ solar models must adopt cubic interpolation in $X$--$Z$ to produce helioseismic results consistent with other published works. We leave further exploration and improvements to \MESA's opacity interpolation methods to future work.

The low temperature opacity tables used in this article will be included in the forthcoming public release of MESA, and can be directly downloaded from \url{http://www.wichita.edu/academics/fairmount_las/physics/Research/opacity.php}. The new Los Alamos OPLIB radiative opacity tables presented in this article will be available at \url{http://aphysics2.lanl.gov/opacity/lanl} and also included in a forthcoming public release of \MESA. Users can also generate opacity tables for their own desired mixtures at \url{http://aphysics2.lanl.gov/opacity/lanl}. We encourage future stellar physics research to experiment with this expanded set of Type-1 Rosseland-mean opacity tables.

\section*{Acknowledgements}

We thank Luis Trivino at Los Alamos National Laboratory for his pivotal help in providing computing resources for this project. We also thank Jason Ferguson and David Alexander for providing the low temperature opacity tables for the MB22 and AAG21 mixtures. This research was partially supported by the National Science Foundation under grant 2154339 and by the U.S. Department of Energy through the Los Alamos National Laboratory. Los Alamos National Laboratory is operated by Triad National Security, LLC, for the National Nuclear Security Administration of U.S. Department of Energy (Contract No. 89233218CNA000001). This research made extensive use of the SAO/NASA Astrophysics Data System (ADS).

\software{
\MESA\ \citep[][\url{https://docs.mesastar.org}]{paxton_2011_aa,paxton_2013_aa,paxton_2015_aa,paxton_2018_aa,paxton_2019_aa,jermyn_2023_aa},
\texttt{MESASDK} 20190830 \citep{mesasdk_linux,mesasdk_macos},
\texttt{matplotlib} \citep{hunter_2007_aa}, 
\texttt{Scipy} \citep{virtanen_2020_aa}, and
\texttt{NumPy} \citep{der_walt_2011_aa}.
         }


\bibliographystyle{aasjournal}

\restartappendixnumbering
\appendix 
Appendices~\ref{appendix.A}, \ref{appendix.B} and \ref{appendix.C} detail the implementation and verification of the new OPLIB tables in \MESA. Appendix~\ref{appendix.A} discusses the use of bicubic splines on the raw OPLIB tables, Appendix~\ref{appendix.B} compares linear and cubic interpolation of opacities across $X$--$Z$ in \MESA, and Appendix~\ref{appendix.C} compares the helioseismic differences between \MESA\ solar models calculated with linear and cubic interpolation of the opacities across the $X$--$Z$ plane.\\

\section{Bicubic splines of OPLIB tables in \MESA} \label{appendix.A}

\begin{figure}[!htb]
    \centering
    \includegraphics[width=5.5in]{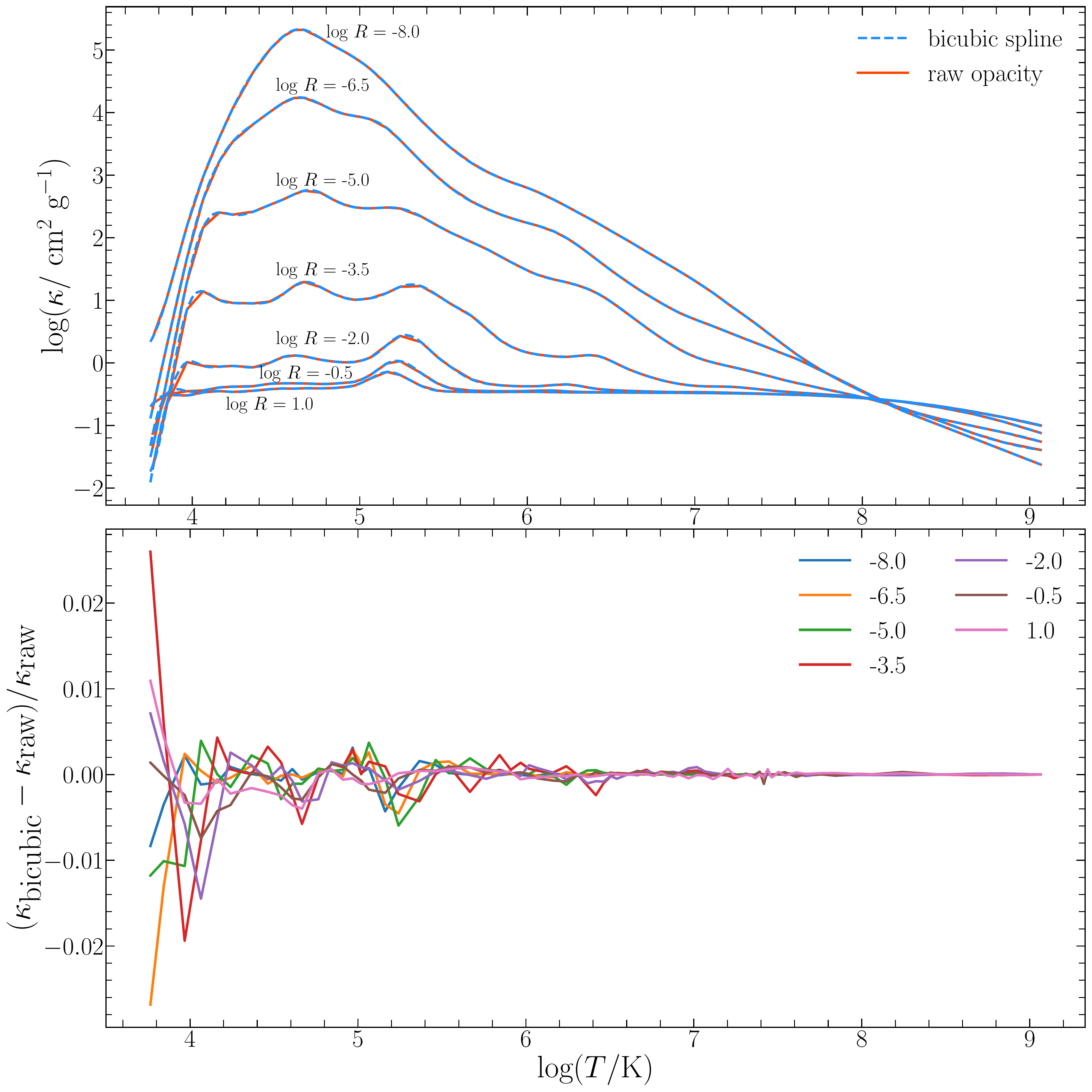}\\
    \caption{Top panel: opacity versus temperature for different log($R$) from a GS98 OPLIB opacity table at $X$=0.7, $Z$=0.02 for different log($R$). The bicubic spline interpolation are overlaid as a dashed curve. Bottom panel: normalized difference between the bicubic spline interpolations and the raw OPLIB tables at various values of log($R$).}
    \label{fig:interp1}
\end{figure}

To ensure smooth opacity derivatives, OPAL and OP opacity tables have historically been run through smoothing and spline-fitting routines, see \citet{seaton_1993_aa}. We generate bicubic spline versions of the raw OPLIB tables using a python Scipy routine. We interpolate the original 74 log($T$/K) $\times$ 39 log($R$) point OPLIB tables spanning 3.764 $\leq$ log($T$/K) $\leq$ 9.065 and $-$8.0 $\leq$ log($R$) $\leq$ 1.5 into evenly spaced 213 log($T$/K) $\times$ 39 log($R$) opacity tables spanning 3.75 $\leq$ log($T$/K) $\leq$ 9.05 with $\Delta T$~$=$~$0.025$, and identical spacing in log($R$). Figure~\ref{fig:interp1} shows the raw OPLIB and bicubic spline fits for a $X$~$=$~$0.7$, $Z$~$=$~$0.02$ mixture along constant log($R$) and their respective normalized differences. Overall, the bicubic spline smoothing of the OPLIB tables results in $\lesssim$~0.5\% changes to the overall opacity, except for log($T$/K)~$\lesssim$~4 where the OPLIB tables are less smooth.

\section{ \MESA\ linear versus cubic interpolation in $X$--$Z$} \label{appendix.B}

\begin{figure*}[!htb]
\centering
\includegraphics[width=6.5in]{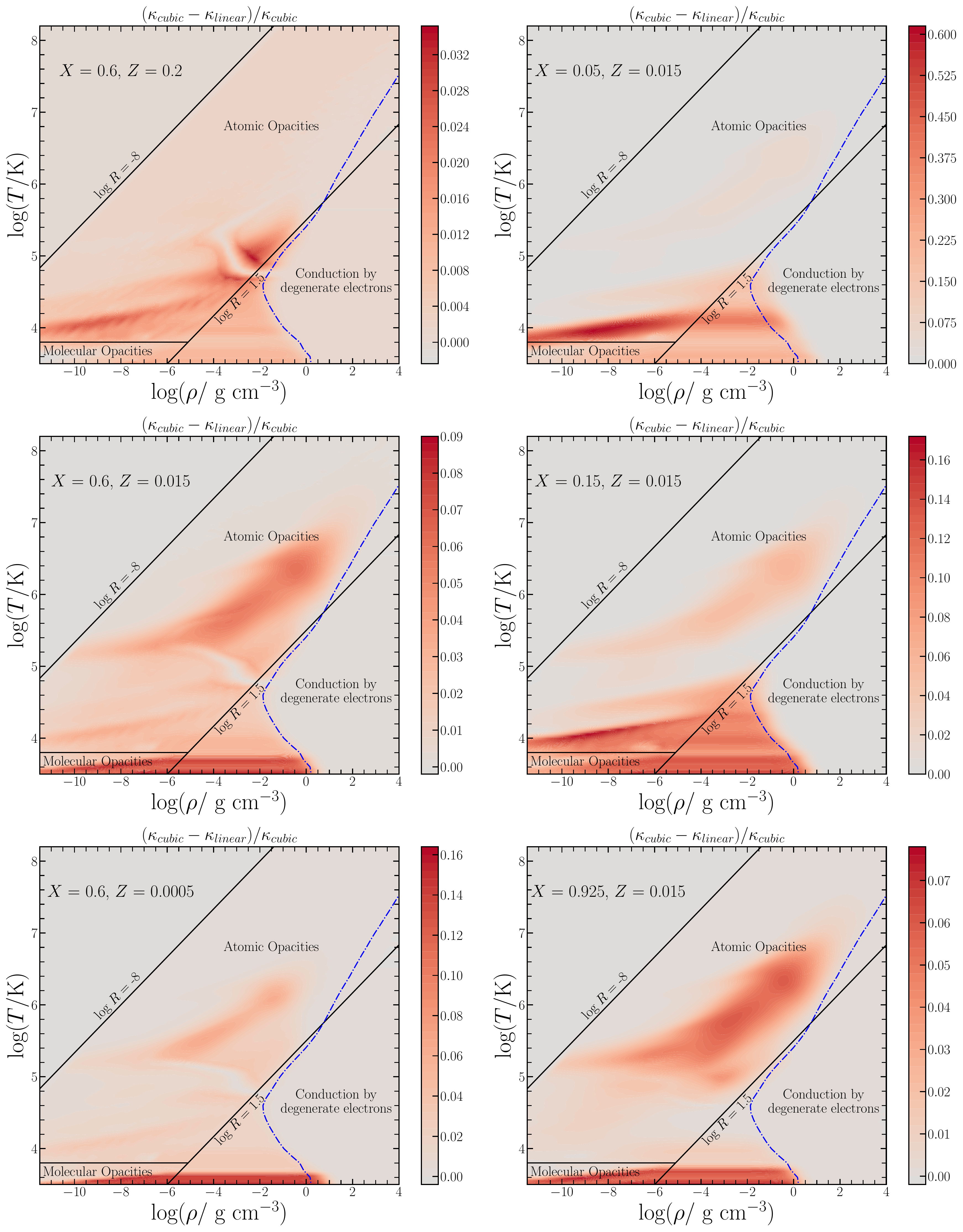} \\
\caption{Relative differences between the 126 OPLIB opacity tables with cubic versus linear interpolation using \citet{grevesse_1998_aa} abundances, generated from \MESA's \texttt{kap} module, for six mixtures. The left column shows mixtures with $X$\,=\,0.6 and varying $Z$, and the right column shows mixtures with $Z$\,=\,0.015 and varying $X$. The OPLIB log($R$)\,=\,-8, 1.5 table boundaries are marked with a solid black line.  The approximate location of the Z-dependent transition to an electron conduction dominated opacity is marked with dot-dash blue curve.}
\label{f.panelE}
\end{figure*}

The default interpolation scheme across the $X$--$Z$ plane in \MESA\ uses linear interpolation:
\begin{itemize}\tightitems
\item[] \code{cubic\_interpolation\_in\_X = .false.} 
\item[] \code{cubic\_interpolation\_in\_Z = .false.}
\end{itemize}
 Figure~\ref{f.panelE} shows the normalized opacity differences between the cubic interpolated OPLIB 126 table set (denoted $\kappa_{cubic}$) and the linear interpolated OPLIB 126 table set (denoted $\kappa_{linear}$) for six mixtures in the $\rho-T$ plane. The six mixtures are chosen to  lie between the available tables in the $X$--$Z$ plane. Overall we find linear interpolation systematically under predicts the opacity by up to 60\%  depending on the metallicity.

\section{\MESA\ solar models with linear versus cubic interpolation in $X$--$Z$} \label{appendix.C}
\begin{figure}[!htb]
\centering
\includegraphics[width=6.5in]{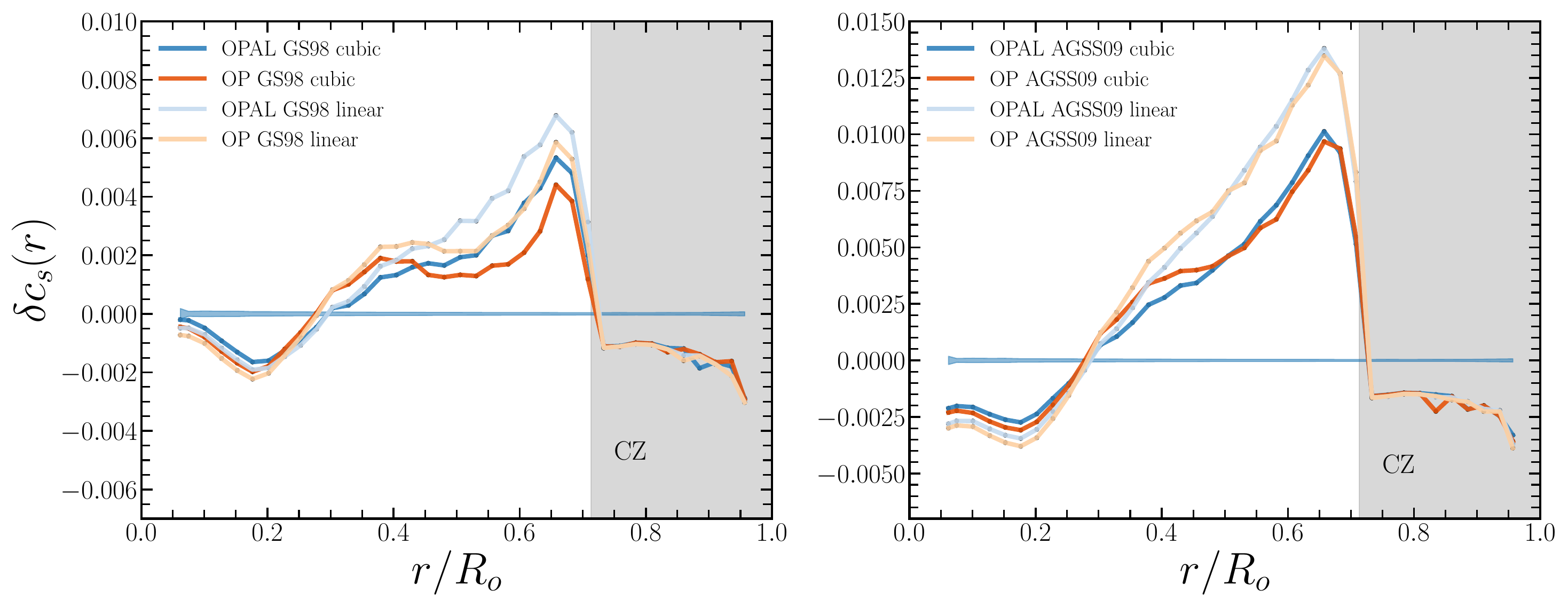} \\
\includegraphics[width=6.5in]{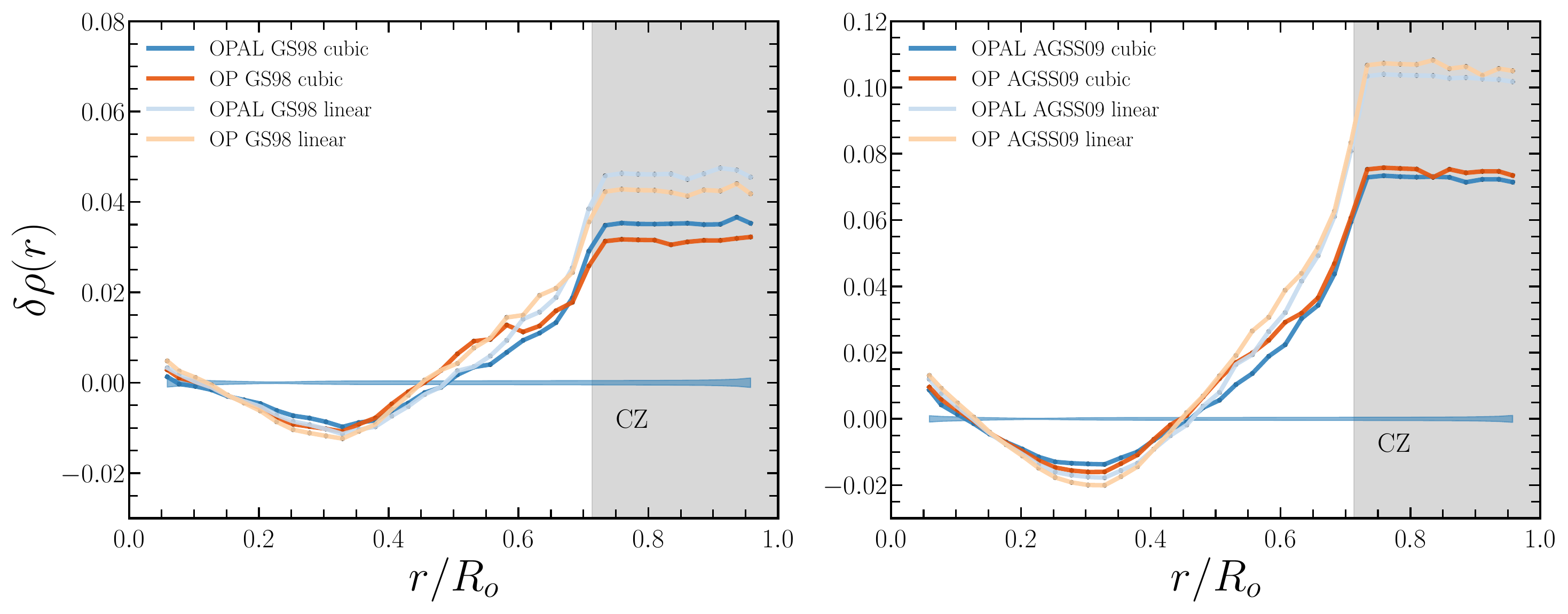} \\
\caption{
Fractional sound speed and density differences, 
$\delta c_{s}$ $=$ $(c_{\rm obs}$ - $c_{s}(r)) /c_{s}(r)$ and $\delta \rho$ $=$ $(\rho_{\rm obs}$ - $\rho(r)) /\rho(r)$, 
between the values predicted by a calibrated \MESA\ standard solar model, $c_{s}(r)$ and $\rho(r)$, 
and the  $c_{\rm obs}$ and $\rho_{\rm obs}$ values inferred from helioseismic data \citep{basu_2009_aa}.
The 1$\sigma$ observational uncertainties are shown as the blue bands at ordinates of zero. 
Black circles mark locations where $\delta c_{s}$ and $\delta \rho$ are evaluated. 
Gray bands shows the convective regions, labeled CZ.
}
\label{f.c_diff_appendix}
\end{figure}

We briefly illustrate the helioseismic differences arising from using linear and cubic interpolation in the $X$--$Z$ plane for standard solar models calibrated with the OP/OPAL opacity tables and GS98/AGSS09 abundances included in \MESA. Figure~\ref{f.c_diff_appendix} shows the relative sound speed and density profiles of solar models calculated with linear and cubic interpolation. The agreement with helioseismic data is systemically worse for solar models that use linear interpolation in $X$--$Z$. The sound speed and density profiles from models that use cubic interpolation are in better agreement with solar model results found in the literature \cite[e.g., ][]{vinyoles_2017_aa,magg_2022_aa} than those run with linear interpolation.

\end{document}